\documentclass[useAMS,usenatbib]{mn2e} 

\usepackage{graphicx} 
\usepackage{natbib} 
\usepackage{times}
\usepackage{amssymb} 
\usepackage{wasysym} 

\catcode`\@=11 
\def\gsim{\ifmmode{\mathrel{\mathpalette\@versim>}} 
   \else{$\mathrel{\mathpalette\@versim>}$}\fi} 
\def\lsim{\ifmmode{\mathrel{\mathpalette\@versim<}} 
   \else{$\mathrel{\mathpalette\@versim<}$}\fi} 
\def\@versim#1#2{\lower 2.9truept \vbox{\baselineskip 0pt \lineskip 
   0.5truept \ialign{$\m@th#1\hfil##\hfil$\crcr#2\crcr\sim\crcr}}} 
\catcode`\@=12 

\newcommand{\beq}{\begin{equation}} 
\newcommand{\eeq}{\end{equation}}

\newcommand{\hii}{H{\sc ii~}} 
\newcommand{\ant}{\rm {\alpha_{nt}}} 
\newcommand{\fth}{f_{\rm th}}

\title[Equipartition magnetic field]{Magnetic fields in nearby normal galaxies: Energy 
equipartition} 

\author[Basu \& Roy]{Aritra Basu$^{1}$\thanks{E-mail: 
   aritra@ncra.tifr.res.in (AB), roy@ncra.tifr.res.in (SR)}, Subhashis Roy$^{1\star}$\\ 
$^{1}$National Center for Radio Astrophysics, TIFR, Pune University Campus, 
Ganeshkhind Road, Pune - 411007.} 

\begin{document} 

\pagerange{\pageref{firstpage}--\pageref{lastpage}} \pubyear{2009} 

\maketitle 

\label{firstpage} 

\begin{abstract} 

We present maps of total magnetic field using `equipartition' assumptions for 
five nearby normal galaxies at sub-kpc spatial resolution. The mean 
magnetic field is found to be $\sim11~\mu$G. The field is strongest near the 
central regions where mean values are $\sim20-25~\mu$G and falls to 
$\sim15~\mu$G in disk and $\sim10~\mu$G in the outer parts.  There is little 
variation in the field strength between arm and interarm regions, such that, in 
the interarms, the field is $\lesssim$20 percent weaker than in the arms. 
There is no indication of variation in magnetic field as one moves along arm or 
interarm after correcting for the radial variation of magnetic field. We also 
studied the energy densities in gaseous and ionized phases of the interstellar 
medium and compared to the energy density in the magnetic field.  The energy 
density in the magnetic field was found to be similar to that of the gas within 
a factor of $\lesssim$2 at sub-kpc scales in the arms, and thus magnetic 
field plays an important role in pressure balance of the interstellar medium. 
Magnetic field energy density is seen to dominate over the kinetic energy
density of gas in the interarm regions and outer parts of the galaxies and
thereby helps in maintaining the large scale ordered fields seen in those
regions.

\end{abstract} 

\begin{keywords} 
galaxies: ISM -- galaxies: magnetic fields -- galaxies: spiral 
-- (ISM:)   cosmic rays -- ISM: general -- radio continuum: ISM. 

\end{keywords} 

\section{Introduction} 
\label{intro} 

Magnetic field strength plays an important role in determining the dynamics and
energetics in a galaxy.  It is believed that the magnetic pressure plays a role
in determining the scale height of the galactic interstellar medium (ISM).
Also, the magnetic field plays an important role in collapse of a gas cloud to
help the star formation activity \citep{elmeg81, crutc99}.  The density and
distribution of cosmic rays depend on magnetic fields. 

It is thought that the seed field, before formation of galaxies, was amplified
by compression during collapse and shearing by a differentially rotating 
disk \citep{beck06}. Dynamo action within the galaxy amplifies and maintains
field strength over galactic life-times of $\sim10^9$ year \citep[see
e.g,][]{moffa78, parke79, moss96, shuku06}. Though the dynamo effect can
amplify the large scale mean magnetic field, magnetohydrodynamic (MHD)
turbulence can amplify the local magnetic field through field line stretching
\citep[][]{batch50, grove03} up to energy equipartition levels. In steady
state, the energy density of magnetic field is close to energy density of the
gas. Gas density is known to fall as a function of galactocentric distance
\citep[see e.g.,][]{leroy08}. Therefore, it is expected that the field strength
will fall as a function of galactocentric distance. 

\begin{table*} 
\centering 
 \caption{The sample galaxies.} 
  \begin{tabular}{@{}lcccccccccc@{}} 
 \hline 
    Name  & Type  &Angular     & $i$& Distance   &  CO & H{\sc i} & FIR&H$\alpha$&Radio\\ 
          &       & size (D$_{25}$)($^\prime$) & ($^\circ$) & (Mpc) & &&$\lambda70\mu$m&&$\lambda20$ cm\\ 
     (1)     &(2)       & (3) & (4) & (5)               & (6) & (7) &(8)& (9)&(10)\\ 
\hline 
NGC 1097    & SBbc     & 9.3$\times$6.3    & 45 & 14.5$^\dagger$     & --&--& SINGS&1.5 m CTIO& VLA$^{a}$ CD array$^4$ \\ 
NGC 4736    & SAab     & 11.2$\times$9.1  & 41 & 4.66$^1$     &HERACLES&THINGS& SINGS&1 m JKT     & Westerbork$^{b}$ SINGS$^5$\\ 
NGC 5055    & SAbc     & 12.6$\times$7.2  & 59 & 9.2$^\dagger$&HERACLES&THINGS& SINGS&2.3 m KPNO&Westerbork SINGS$^5$\\ 
NGC 5236    & SABc     & 11.2$\times$11 & 24 & 4.51$^2$ &NRAO 12 m&THINGS& SINGS&0.9 m CTIO& VLA CD array$^6$\\ 
NGC 6946    & SABcd    & 11.5$\times$9.8  & 33 & 6.8$^3$       &HERACLES&THINGS& SINGS&2 m KPNO& VLA C+D array$^7$\\ 
\hline 
\end{tabular} 

In column 3 D$_{25}$ refers to the optical diameter measured at the 25
magnitude arcsec$^{-2}$ contour from \cite{vauco91}.  Column 4 gives the
inclination angle ($i$) defined such that $0^\circ$ is face-on. Distances in
column 5 are taken from: $^1$ \cite{karac03}, $^2$ \cite{karac02}, $^3$
\cite{karac00} and the NED $^\dagger$. Columns 6 and 7 lists the data used to
trace the molecular and atomic gas respectively which were used to estimate the
gas density. Column 9 lists the sources of H$\alpha$ maps used to estimate the
energy density of ionized gas in Section 4.2. Column 10 lists the sources of
archival data at $\lambda20$ cm waveband: $^4$ VLA archival data using CD array
configuration (project code: AW237), $^5$ \citet{braun07}, $^6$ VLA archival
data using CD array (project code: AS325), $^7$ VLA archival map by combining
data from C and D array \citep{beck07}.\\ $^a$ The Very Large Array (VLA) is
operated by the NRAO. The NRAO is a facility of the National Science Foundation
operated under cooperative agreement by Associated Universities, Inc.\\ $^b$
The Westerbork Synthesis Radio Telescope (WSRT) is operated by the Netherlands
Foundation for Research in Astronomy (NFRA) with financial support from the
Netherlands Organization for scientific research (NWO). 
\label{sampletab} 
\end{table*} 

Observationally, the magnetic field ($B$) can be traced by polarization studies
at various wavebands, e.g., Faraday rotation and synchrotron radiation
polarization in radio, polarization of starlight in optical and polarized dust
emission in infrared.  Zeeman splitting of spectral lines can be used to
estimate the local magnetic field. Intensity of the synchrotron emission at
radio wavelengths can provide estimates of $B$ though assumptions of
`equipartition' of energy between cosmic ray particles and magnetic field. 

Faraday rotation can probe the line-of-sight averaged magnetic field
($B_\parallel$). However, this method uses polarized radio emission that may
not be seen from a large fraction of a galaxy due to Faraday and/or beam
depolarization \citep{sokol98}.  Zeeman splitting can directly measure
$B_\parallel$, but it is highly susceptible to high localized magnetic field.
Moreover, its detection is difficult in external galaxies. Estimation of $B$ in
the sky-plane using polarization of starlight or dust emission depends highly
on several geometrical and physical parameters \citep{zweib97}.  Synchrotron
emission is seen from large fraction of a galaxy and under the condition of
`equipartition', it provides a measure of total magnetic field.  At low
frequencies (0.33 GHz), more than 95 percent of the emission is synchrotron in
origin \citep{basu12a} therefore low radio frequency total intensity images can
be used to determine $B$ in galaxies. 

This method has been used to determine $B$ in some of the nearby star forming
galaxies. In M51, \citet{fletc11} found generally stronger fields of
$\sim20-25~\mu$G in the spiral arms and $\sim15-20~\mu$G in the interarm
regions.  In this case, $B$ was determined using total intensity map at
$\lambda6$ cm assuming a constant spectral index for thermal and synchrotron
emission. This might introduce errors in the results as the nonthermal spectral
index steepens from center to edge \citep{basu12a}.  In NGC 253, the field was
found to be $\sim20~\mu$G towards the center and fell to $\sim8~\mu$G towards
the edge \citep{heese09}.  In M82, the total field was found to be
$\sim80~\mu$G in the center and $\sim20-30~\mu$G in the synchrotron emitting
halo \citep{adebh12}.  However, these galaxies are known starbursts, the
magnetic field in the disk could be significantly affected by mixing of
magnetic field from other parts of the galaxy through galactic fountain
\citep[][]{shapi76, bregm80, norma89, heald12}. 

To measure magnetic field and to compare its energy density with that in gas at
high spatial resolution of $\lesssim1$ kpc we have observed five nearly face-on
normal galaxies, namely, NGC 1097, NGC 4736, NGC 5055, NGC 5236 (M83) and NGC
6946.  In Section 2, magnetic field strengths in these galaxies are determined
using total intensity synchrotron emission at 0.33 GHz. In Section 3, we
present the magnetic field maps and results. We discuss our results and compare
the magnetic field energy with kinetic energies in turbulent gas in various
phases of the ISM. Our results are summarized in Section 5.

\section{Data analysis} 

The galaxies studied here were chosen from \citet{basu12a} where a thorough
separation of thermal emission from the total emission was done at 0.33 GHz
($\lambda90$ cm) and near 1.4 GHz ($\lambda20$ cm) using H$\alpha$ as the
tracer after correcting for dust absorption \citep{tabat07}. Our sample
comprises of NGC 1097, NGC 4736, NGC 5055, NGC 5236 and NGC 6946.  The
nonthermal spectral index ($\ant$) used to compute the equipartition magnetic
field, was estimated between the above mentioned frequencies.  The data sources
for the sample galaxies are listed in Table 1. The 0.33 GHz observations were
made using the Giant Meterwave Radio Telescope (GMRT). We broadly classified
our studies between arm and interarm regions, i.e, regions of high and low gas
density, identified from the H$\alpha$ images for each galaxy.

Due to poorer resolution of the far infrared maps used for determining
absorption correction of the H$\alpha$ emission, the overall resolution of the
nonthermal emission maps was only 40 arcsec. To improve the resolution of
nonthermal emission we used $\lambda24$-$\mu$m emission from dust as a tracer
of thermal emission \citep{murph08}. The {\it Spitzer} MIPS $\lambda24$-$\mu$m
maps have a resolution of 6 arcsec, better than the resolution of the radio
maps.  The resolution of the nonthermal maps are determined by the lowest
resolution radio maps and subsequently the $\lambda24$-$\mu$m maps were
convolved to it.  However, $\lambda24$-$\mu$m emission from dust is not a
direct tracer of thermal emission, and in certain cases show differences with
measurements made from using a direct tracer like H$\alpha$ \citep{perez06,
calze05}. Moreover, the $\lambda24$-$\mu$m emission arises not only from
dust grains heated by ultra violet (UV) photons, but also from heating of
diffuse cirrus clouds by the interstellar radiation field and also from old
stars, mostly from the central regions. This could lead to overestimation of
thermal emission in such regions. To avoid this shortcoming, and to ensure
that both the methods give identical results at the resolution of the
absorption corrected H$\alpha$ emission, we corrected the thermal
fraction\footnote{Thermal fraction is defined as: $\fth = {S_{\nu,\rm
th}}/{S_{\nu,\rm tot}}$, where, $S_{\nu,\rm tot}$ and $S_{\nu,\rm th}$ are the
flux densities of the total and thermal radio emission respectively at a radio
frequency $\nu$.} determined from $\lambda24$-$\mu$m to the thermal fraction
from H$\alpha$ in the method described below.   All the maps were brought to
the same pixel size (3 arcsec) and aligned to a common coordinate system. All
the pixels with signal-to-noise ratio more than 4 were considered for this
analysis.

\begin{figure} 
\begin{center}
\includegraphics[height=7cm,width=7cm]{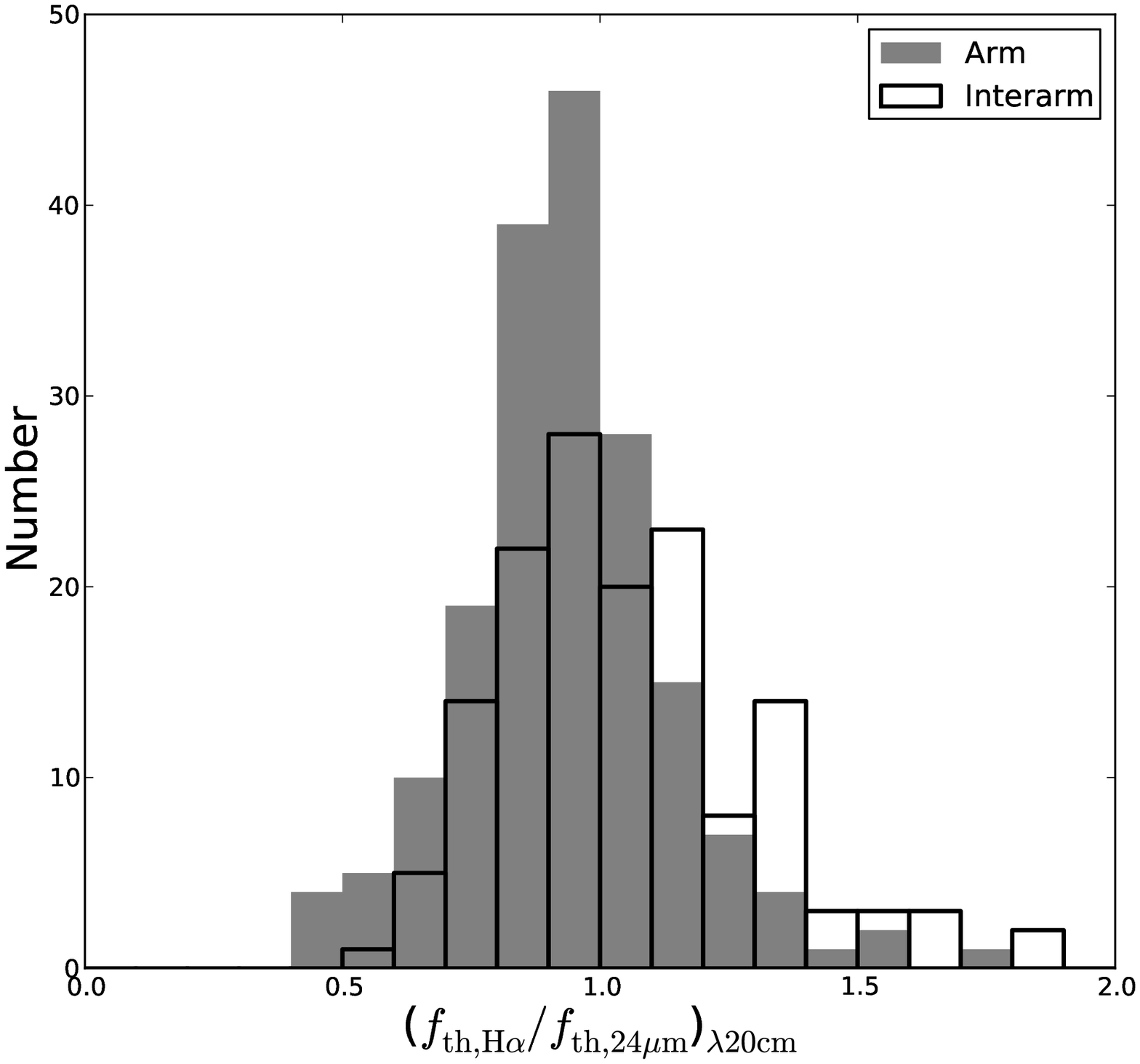} 
\includegraphics[height=7cm,width=7cm]{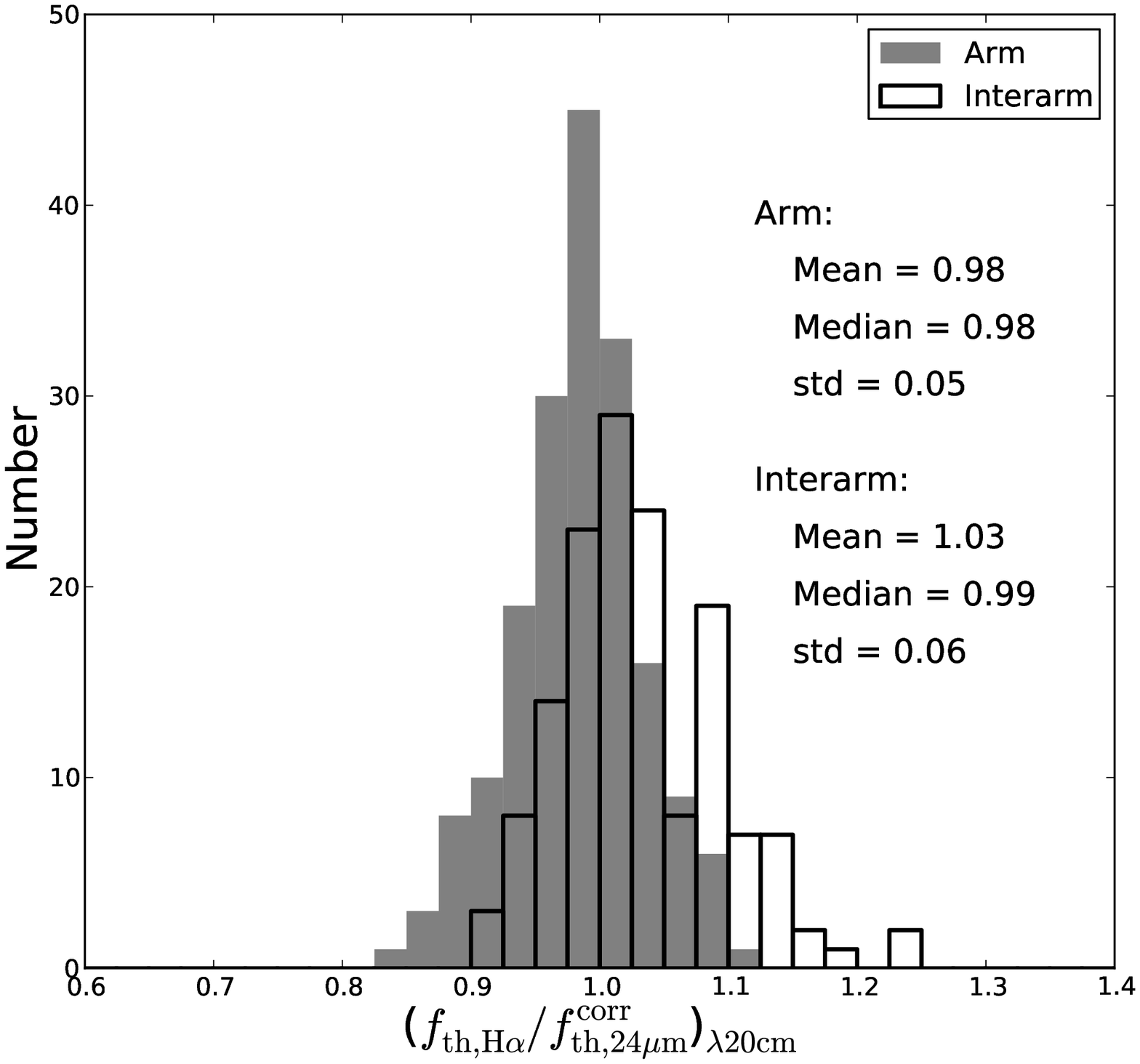} 
\caption{Top: histogram of the correction factor ($f_{\rm th, H\alpha}/f_{\rm
th,24\mu m}$) after normalizing with median values of the ratio for each
galaxy; bottom: distribution of $f_{\rm th, H\alpha}/f_{\rm th,24\mu m}^{\rm
corr}$ for all the galaxies at 1.4 GHz determined within 40 arcsec regions.
The grey and unfilled histograms are for arm and interarm regions.}
\end{center}
\label{fthfig}  
\end{figure}

\begin{table} 
\centering 
 \caption{The resolution of available radio maps in units of arcsec$^2$.} 
  \begin{tabular}{@{}lcccc@{}} 
 \hline 
    Name  & $\lambda90$ cm  & $\lambda20$ cm   & nonthermal & linear   \\
          &       &  &  maps & scale (kpc) \\
\hline 
NGC 1097    &  16$\times$11    & 40$\times$30      & 40$\times$40 & 2.80 \\ 
NGC 4736    &  13$\times$12    & 19$\times$12.5    & 20$\times$20 & 0.45\\
NGC 5055    &  17$\times$10    & 18.5$\times$12.5  & 20$\times$20 & 0.90\\
NGC 5236    &  16$\times$12    & 26$\times$14      & 26$\times$14 & 0.55\\
NGC 6946    &  12$\times$11    & 15$\times$15      & 15$\times$15 & 0.50\\
\hline 
\end{tabular} 

\label{restab} 
\end{table}

\begin{table} 
\centering 
 \caption{Mean values of the thermal fractions at $\lambda20$ cm determined
using H$\alpha$ method (column 2) and $\lambda24\mu$m method (column 3).}
\begin{tabular}{@{}lccc@{}} 
 \hline 
    Name  & $\langle f_{\rm th,H\alpha}\rangle$  & $\langle f_{\rm th,24\mu m}\rangle$   & $\left\langle f_{\rm th,H\alpha}/f_{\rm th,24\mu m} \right\rangle$    \\
          &   ($\%$)    & ($\%$) &     \\
\hline 
NGC 1097    &  5$\pm$3    & 7$\pm$2        & 0.75$\pm$0.3 \\ 
NGC 4736    &  7$\pm$2    & 10$\pm$2       & 0.67$\pm$0.2 \\
NGC 5055    & 10$\pm$3    & 11$\pm$2       & 0.87$\pm$0.3 \\
NGC 5236    &  7$\pm$2    & 8.5$\pm$3      & 0.73$\pm$0.3 \\
NGC 6946    & 10$\pm$3    & 8.5$\pm$2      & 1.18$\pm$0.4 \\
\hline 
\end{tabular} 
\label{fthtab} 
\end{table}

In step (i), the convolved $\lambda24$-$\mu$m emission was used to estimate the
thermal emission using Equation 10 in \citet{murph08} at a resolution given in
Table~\ref{restab}. We then estimated the thermal fraction at each pixel
of the map. In step (ii), the maps made in step (i) are convolved to a
resolution of 40 arcsec. At this resolution, the thermal fraction maps made
from $\lambda24~\mu$m ($f_{\rm th, 24\mu m}$) must match the corresponding
thermal fraction maps made from H$\alpha$ ($f_{\rm th, H\alpha}$). Therefore,
in step (iii) we divided the thermal fraction maps made from H$\alpha$ by the
maps made in step (ii). The ratio is expected to be $\sim1$. However, note
that, Equation 10 in \citet{murph08} uses the calibration for the galaxy M51 to
scale dust emission at $\lambda24~\mu$m to trace thermal emission. This is
known to vary between galaxies and may have systematic offsets between $f_{\rm
th, H\alpha}$ and $f_{\rm th, 24\mu m}$. Table~\ref{fthtab} shows the thermal
fraction determined using the H$\alpha$- and $\lambda24\mu$m-method (columns 2
and 3 respectively). The two methods match well within $\sim30$ percent of each
other.  In step (iv), the correction factor to scale the $f_{\rm th, 24\mu m}$
for each pixel was determined within beam of 40 arcsec from the ratio map
determined in step (iii). This correction factor for each pixel was
multiplied with map (i) to obtain the corrected thermal fraction map ($f_{\rm
th, 24\mu m}^{\rm corr}$). The mean correction factor for each of the galaxies
are listed in column 4 of Table~\ref{fthtab}. The correction factor would take
care of the systematic calibration-offsets between galaxies. The resultant
maps provide us with thermal fraction of the galaxies with a resolution better
than 40 arcsec.  Fig.~1, top panel, shows the histogram plot of the ratio
$f_{\rm th, H\alpha}/f_{\rm th, 24\mu m}$ determined within regions of 40
arcsec for all the galaxies at $\lambda20$ cm.  The ratio has been normalized
by the mean values of each galaxy to account for the systematic offset between
galaxies.  The grey and unfilled histograms are for arm and interarm regions
respectively.  For $\sim65$ percent of the regions, the ratio is seen to be
smaller than unity suggesting $\lambda24$-$\mu$m emission to be higher than the
star formation rate.  Fig.~1, bottom panel, compares the thermal fraction
determined within 40 arcsec regions using H$\alpha$- and corrected
$\lambda24\mu$m-method at $\lambda20$ cm.  Although, after correction, $f_{\rm
th, 24\mu m}^{\rm corr}$ agrees with $f_{\rm th, H\alpha}$ within $\sim10$
percent, there is significant spread.  However, to the first order, when
compared to $f_{\rm th, H\alpha}$, $f_{\rm th,24\mu m}^{\rm corr}$ has
significantly less spread and systematic offset than $f_{\rm th,24\mu m}$.
Thermal emission was estimated using $f_{\rm th,24\mu m}^{\rm corr}$ and was
subtracted from the total emission to obtain the nonthermal emission. The
resolution of the nonthermal maps thus obtained using this method are given in
Table~\ref{restab}.

\begin{figure*} 
\includegraphics[width=7.5cm,height=7.5cm]{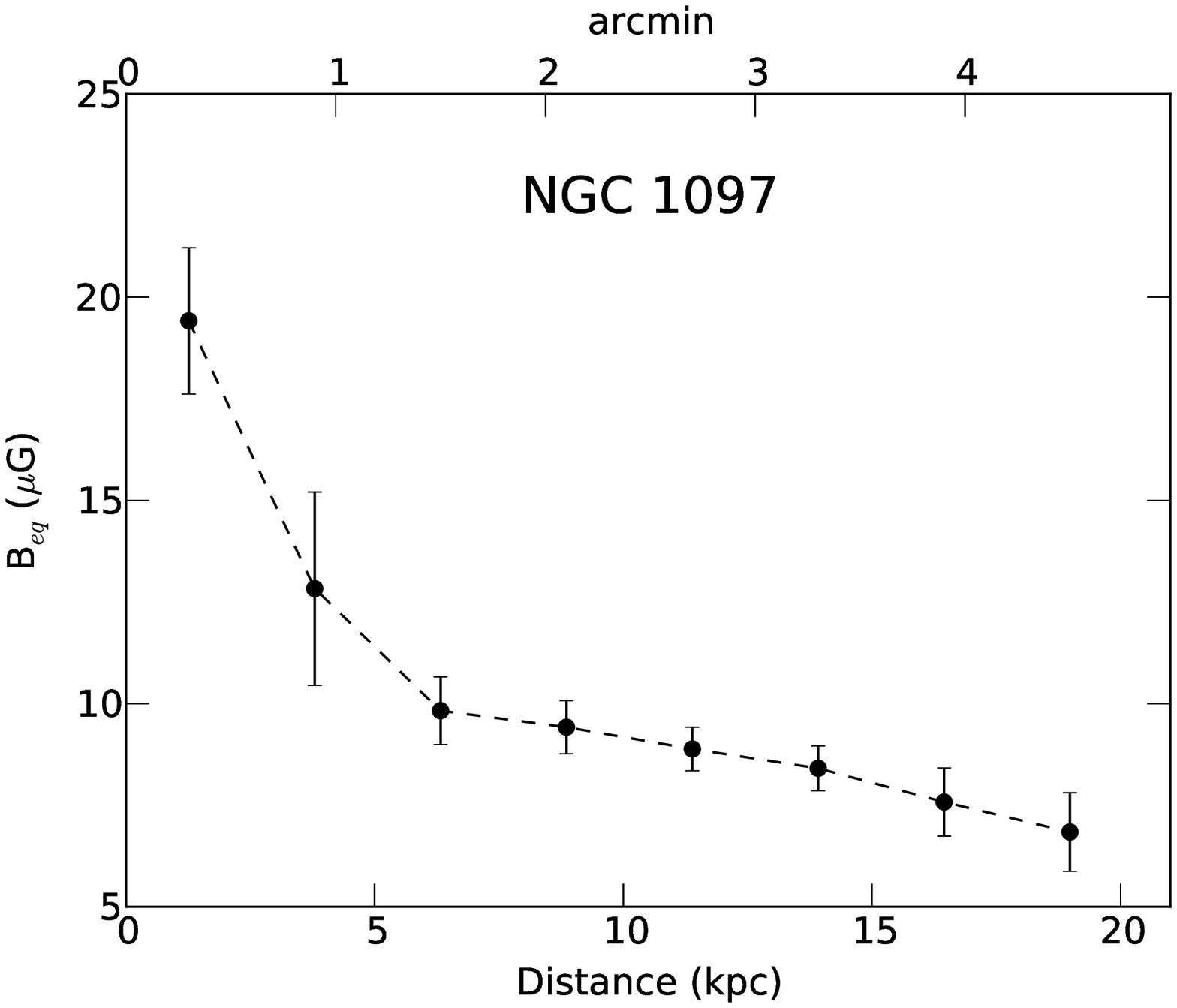} 
\includegraphics[width=7.5cm,height=7.5cm]{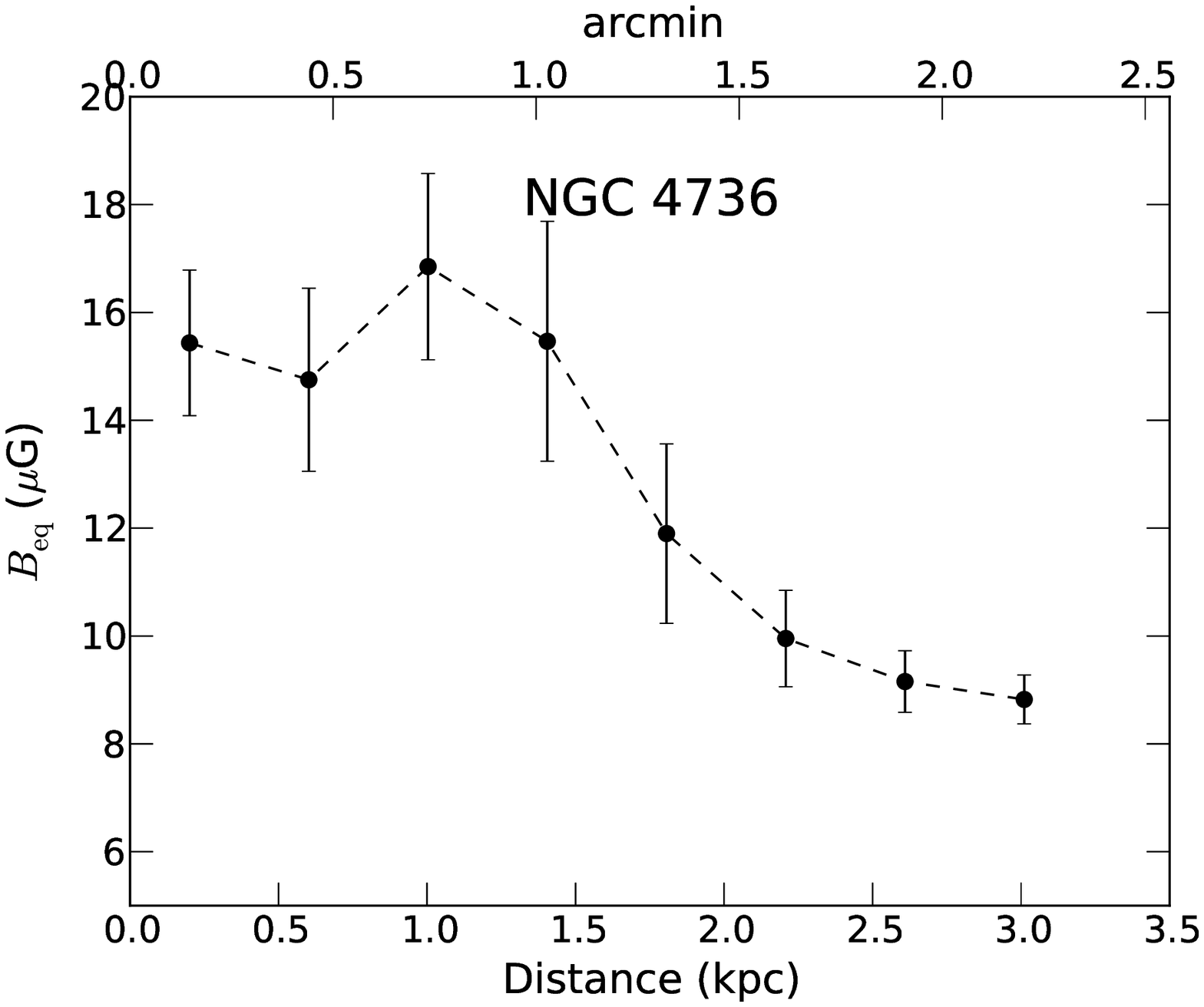} 
\includegraphics[width=7.5cm,height=7.5cm]{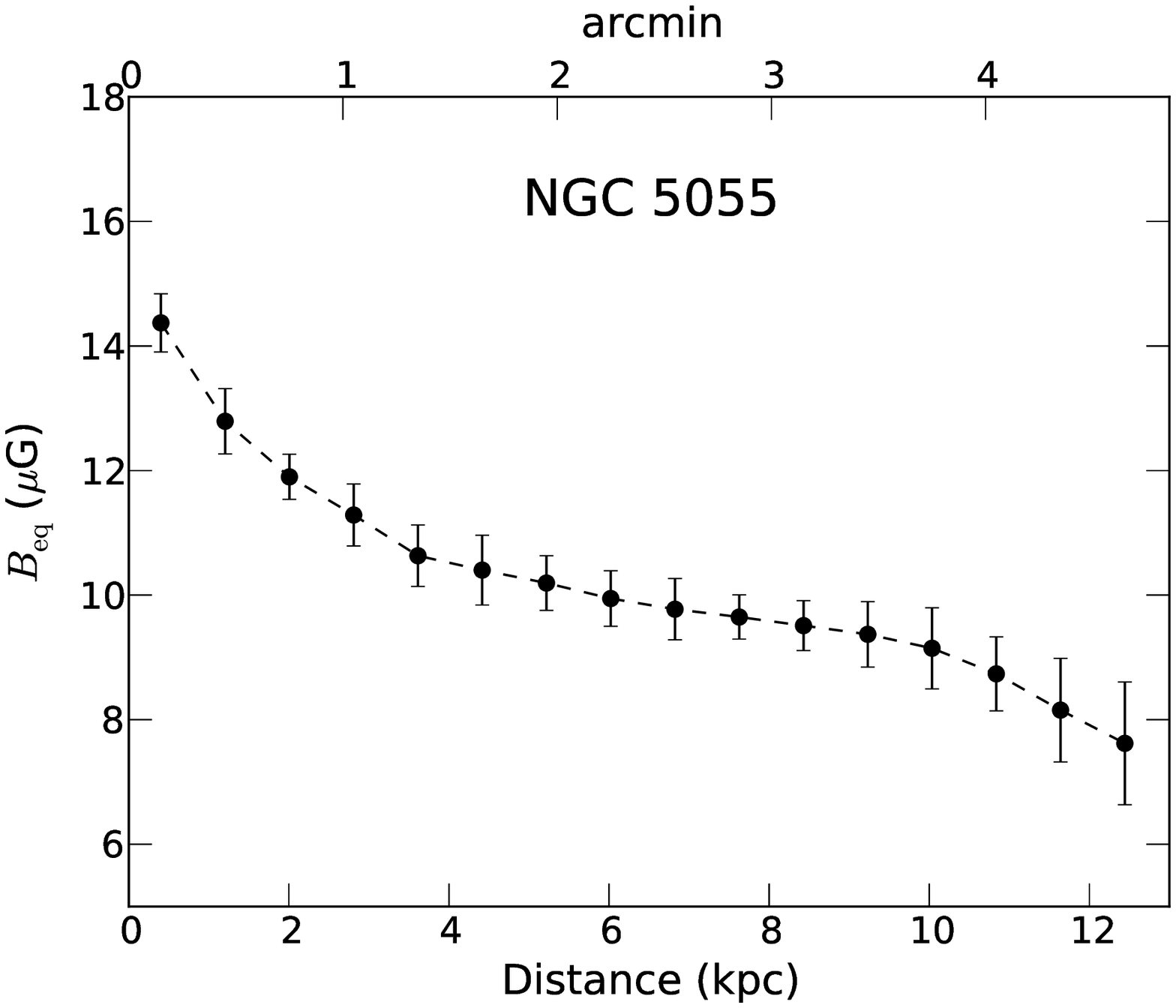} 
\includegraphics[width=7.5cm,height=7.5cm]{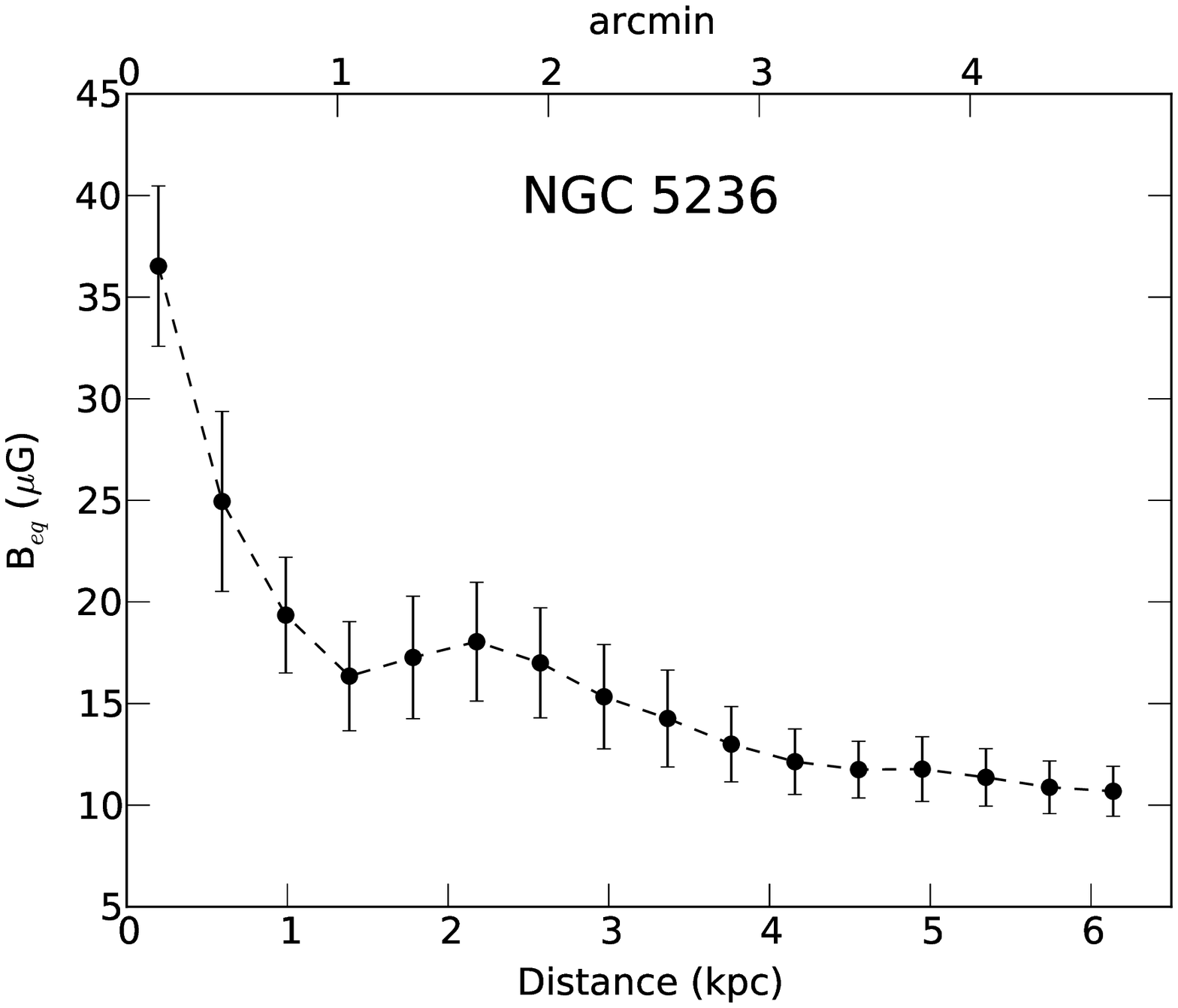} 
\includegraphics[width=7.5cm,height=7.5cm]{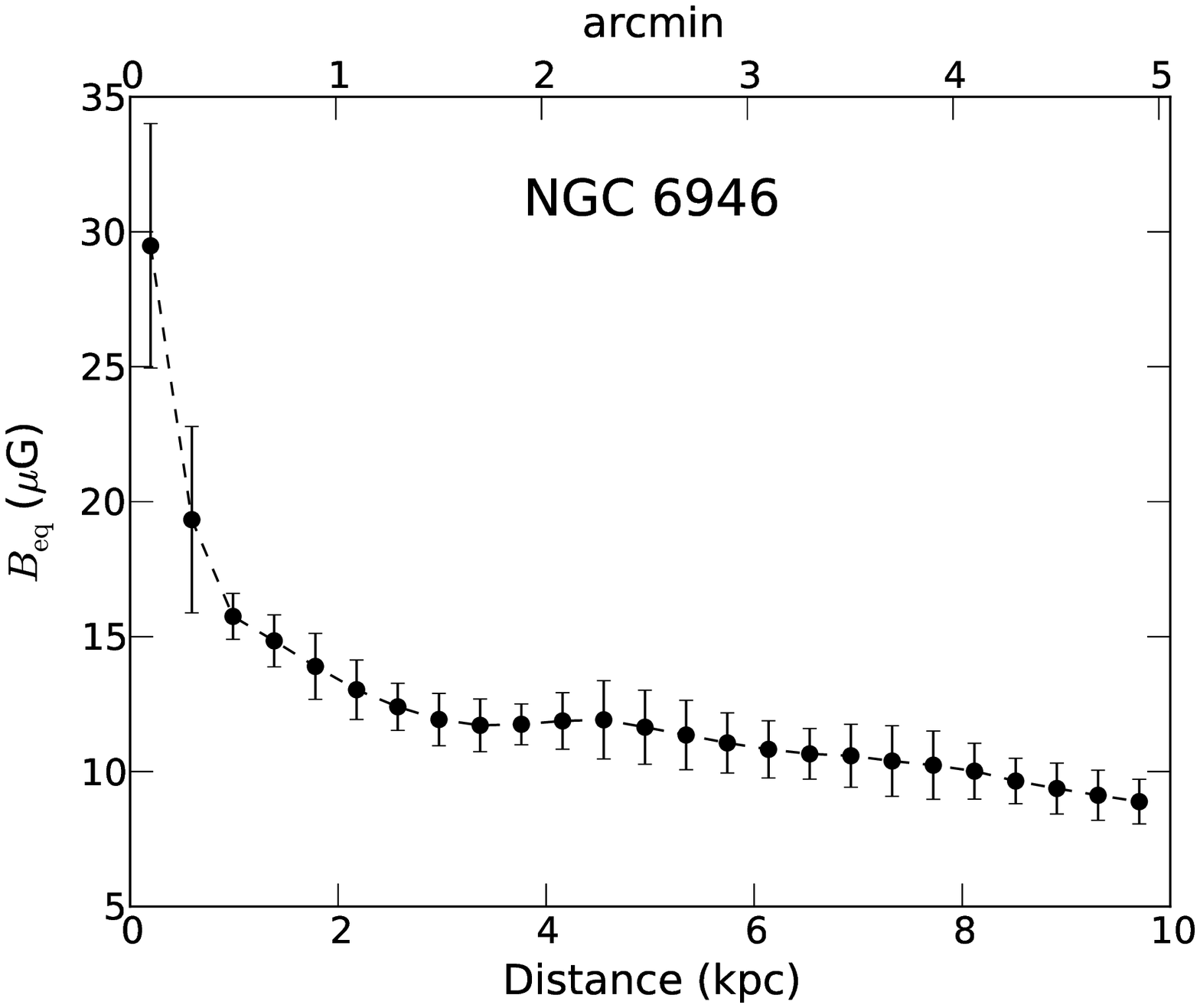} 
\caption{The total equipartition magnetic field maps (in $\mu$G) for the
galaxies NGC 1097, NGC 4736, NGC 5055, NGC 5236 and NGC 6946. The maps have
angular resolution of 40$\times$40 arcsec$^2$, 20$\times$20 arcsec$^2$,
20$\times$20 arcsec$^2$, 26$\times$14 arcsec$^2$ and 15$\times$15 arcsec$^2$
respectively (shown in the bottom left corner). The errors in the central
(red) regions was found to be $\sim2\%$, in the disk (green regions)
$\sim5-10\%$ and in the outer parts (blue regions) $\sim15-20\%$.  Overlaid are
the 0.33 GHz contours from \citet{basu12a}.} 
\label{bmaps} 
\end{figure*}

For the spatially resolved study of the energy densities in the ISM, we used
data from THINGS \citep{walte08} to trace H{\sc i} surface mass density, and
from HERACLES \citep{leroy09} and NRAO 12-m telescope \citep{crost02} to trace
H$_2$ surface mass density (see Appendix A for details).  All the maps for a
galaxy was convolved to a common resolution of the nonthermal maps (see
Table~\ref{restab}) and re-gridded to common pixel size of 3 arcsec. They were
then aligned to the same coordinate system for further analysis.

\begin{figure*}
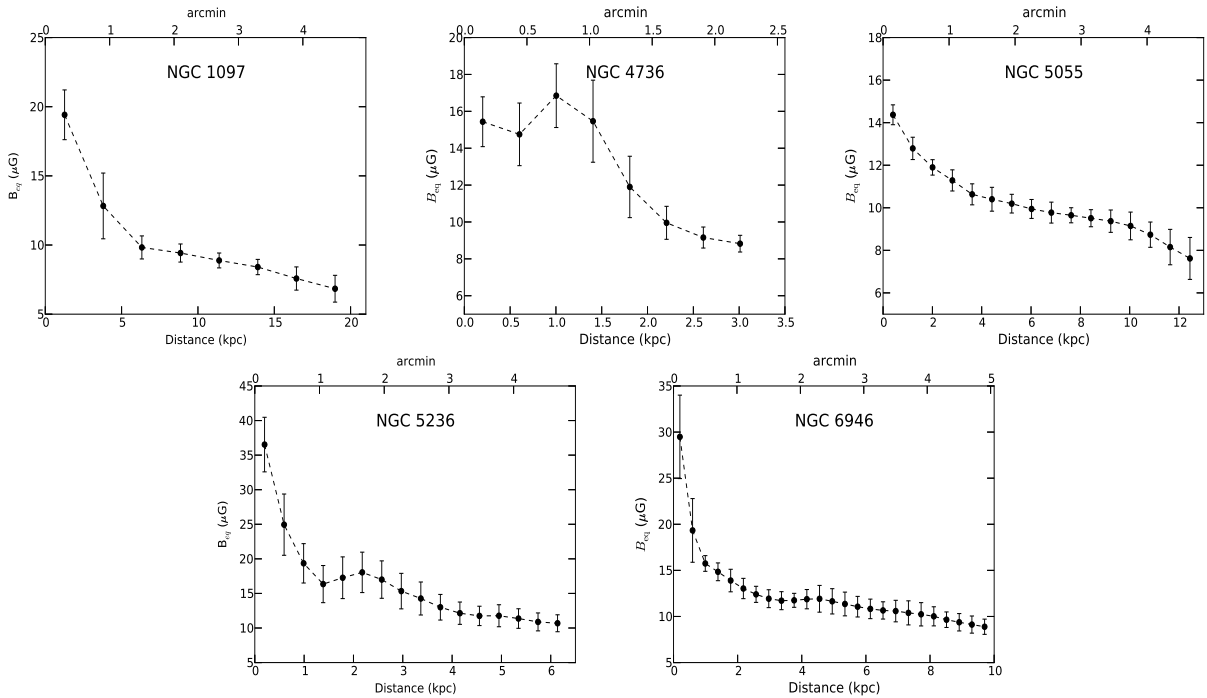
 
\includegraphics[width=5.5cm,height=4.6cm]{ngc1097_brev_lb.eps} 
\includegraphics[width=5.5cm,height=4.6cm]{ngc4736_brev_lb.eps} 
\includegraphics[width=5.5cm,height=4.6cm]{ngc5055_brev_lb.eps} 
\includegraphics[width=5.5cm,height=4.6cm]{ngc5236_brev_lb.eps} 
\includegraphics[width=5.5cm,height=4.6cm]{ngc6946_brev_lb.eps} 
\caption{Variation of the total equipartition magnetic field 
strength as a function of galactocentric distance.} 
\label{bradial} 
\end{figure*}

\subsection{Total magnetic field} \label{totmag} 

From basic synchrotron theory, and assuming energy equipartition between cosmic
ray particles and the magnetic field, the total field strength could be
estimated \citep[see e.g.,][]{pacho70, miley80, longa11}. However, the limits
of integration ($\nu_{\rm min}$ to $\nu_{\rm max}$) over the synchrotron
spectrum to estimate the total energy in cosmic ray electrons (CRe) depends on
the magnetic field \citep{beck05}. This was ignored while minimizing the total
energy in magnetic field and cosmic ray particles to derive the `classical'
equipartition formula \citep[Equation 2 of][]{miley80}.  This gives rise to
overestimation of the field in regions of steep nonthermal spectral index
($\ant > 0.7$, defined as $S_\nu \propto \nu^{-\ant}$). 

We used the `revised' equipartition formula given in Equation 3 of 
\citet{beck05} to produce total magnetic field maps, where the equipartition 
field strength ($B_{\rm eq}$) is given as, 
\begin{equation} 
B_{\rm eq} = \left\{ 4\pi (K_0+1) E_p^{1-2\ant}\frac{f(\ant)}{c_4(i)}\frac{I_\nu \nu^{\ant}}{l}\right\}^{1/(\ant+3)}. 
\end{equation} 
\label{beq} 
Here, $E_p$ is the rest mass energy of protons, $I_\nu$ is the nonthermal
intensity at frequency $\nu$, $l$ is the path-length through the synchrotron
emitting region. $K_0$ is the ratio of number density of relativistic protons
and electrons, $c_4(i)$ is a constant depending on the inclination angle of the
magnetic field. $f(\ant)$ is a function of $\ant$ such that, $f(\ant) =
(2\ant+1)/\left[2(2\ant-1)c_2(\ant) c_1^{\ant}\right]$, and $c_1$, $c_2$ are
constants defined in Appendix of \citet{beck05}. 

We assume $K_0$, the ratio of number densities of relativistic protons ($n_{\rm
CRp}$) and electrons ($n_{\rm CRe}$), such that $K_0 = n_{\rm CRp} / n_{\rm
CRe} \simeq 100$. The path-length travelled through the source ($l$) is taken
to be 2 kpc and corrected for the inclination.  This could in principle be a
function of galactocentric distance ($r$, i.e, $l\equiv l(r)$) depending on the
shape of synchrotron emitting halo perpendicular to plane of galaxy disk.   The
scale height of the synchrotron emitting halo depends on the synchrotron
lifetime ($\tau_{\rm syn}$), and is expected to be uniform along the extent of
the disk except perhaps near the central parts of the galaxies ($\sim1$ kpc) or
in high density regions. Also, $B(r) \propto l(r)^{-1/(\ant+3)}$ (see Eq. 1)
shows weak dependence of magnetic field on $l(r)$.  Therefore, we assume that
the path-length through the source to be constant ($l \equiv l(r) = l_0 \simeq
2$ kpc). The magnetic field thus estimated by us can be scaled by
$[2\times10^{-2}(K_0+1)/l]^{1/(\ant+3)}$ due to the assumption of $K_0 =100$
and $l_0= 2$ kpc. 

The `revised' equipartition formula in Eq. 1 diverges for $\ant \leq 0.5$.
Thus for regions where $\ant$ was found to be less than 0.55, mostly in the
center and inner arms of NGC 5236 and some parts in the ring of NGC 4736, we
used a spectral index of 0.55 to avoid sudden rise in the total field strength.
Such regions have high gas densities and perhaps dominated by ionization or
bremsstrahlung losses giving rise to flatter $\ant$ \citep[see][]{longa11}.  As
a result, the magnetic field strength is overestimated in such regions
\citep{lacki13}. The regions of steep spectral index ($\ant > 1$) towards the
outer parts of the galaxies arises due to dominant energy losses of CRe. 
Thus the energy spectral index between CRe and cosmic ray protons changes,
which is assumed to be constant and the same between protons and electron in
the equipartition formula. We have therefore set $\ant$ as 1 for such regions.
This gives $\sim$6--10 percent lower field strength as compared to steeper
$\ant$. 

\begin{table} 
 \caption{Mean equipartition magnetic fields.} 
\centering
  \begin{tabular}{@{}lccc@{}} 
 \hline 
    Name  & $\langle B_{\rm eq} \rangle$  & $\langle B_{\rm eq} \rangle_{\rm arm}$  & $\langle B_{\rm eq} \rangle_{\rm interarm}$  \\
&($\mu$G)&($\mu$G)&($\mu$G)\\
\hline 
NGC 1097    &   9.0$\pm$2.0    &  9.6$\pm$2.2      & 8.9$\pm$2.3  \\ 
NGC 4736    &   9.3$\pm$2.1    & 16.6$\pm$2.4      &  9.5$\pm$0.7 \\
NGC 5055    &   9.5$\pm$1.1    & 10.2$\pm$0.6      & 9.9$\pm$0.6 \\
NGC 5236    &  12.2$\pm$2.5    & 14.6$\pm$3.0      & 12.0$\pm$2.4 \\
NGC 6946    &  10.7$\pm$1.8    & 12.3$\pm$1.8      & 11.2$\pm$1.2 \\
\hline 
\end{tabular}\\ 
{Note: The mean magnetic field strength for the galaxies were computed
including the low surface brightness diffuse emission and is therefore 
less than the mean values in arm and interarm regions.}
\label{bmeantab} 
\end{table}

\section{Results}

The estimated `equipartition' magnetic field strength for the five galaxies,
using Eq. 1, are shown in Figure~\ref{bmaps}. The resolution of the maps for
each galaxy are tabulated in Table~\ref{restab} and is shown in the lower left
corner of each image.  Overlaid are the 0.33 GHz contour maps of the galaxies
from \citet{basu12a}. The galaxy integrated mean values of magnetic field,
$\langle B_{\rm eq}\rangle$, are found to be $9.0\pm2.0~\mu$G,
$9.3\pm2.1~\mu$G, $9.5\pm1.1~\mu$G, $12.2\pm3.0~\mu$G and $10.7\pm1.8~\mu$G
for NGC 1097, NGC 4736, NGC 5055, NGC 5236 and NGC 6946 respectively (see
Table~\ref{bmeantab}). 

Figure~\ref{bradial} shows $B_{\rm eq}$ as a function of galactocentric 
distance ($r$) estimated by azimuthal averaging over annuli of one beam width. 
The field strength are found to be strongest near the central regions with 
$\langle B_{\rm eq}\rangle\sim 20-25~\mu$G. In the disk, $\langle B_{\rm 
eq}\rangle$ falls to $\sim 15~\mu$G and $\sim10~\mu$G in the outer parts of the 
galaxy.  That is, in most of the cases it is seen that the magnetic field fall 
by $\sim$40--50\% from the center to the edge, similar to what is seen for the 
Milky Way \citep{beck96a}. 

The errors in the magnetic field strength was estimated using Monte-Carlo 
method, wherein $\sim10^4$ random flux density samples were generated assuming 
Gaussian distribution of error in source flux densities at each frequency. 
These were used to determine the distribution of $B_{\rm eq}$. For high 
signal-to-noise regions ($\gsim10\sigma$, i.e towards the inner parts of the 
galaxies ) the distribution of $B_{\rm eq}$ can be modelled as Gaussian. 
However, for regions with lower signal-to-noise ($\lesssim5\sigma$, i.e, in the 
outer parts of the galaxies) the distribution has a tail. The error in the 
total field strength was found to be $\sim2\%$ towards the central regions 
(corresponding to red regions in Figure~\ref{bmaps}), $\sim5-10\%$ in the disk 
(green regions in Figure~\ref{bmaps}) and  $\sim15-20\%$ in the outer parts 
(blue regions in Figure~\ref{bmaps}).

We compared the magnetic field determined using the revised and the classical 
formula. In the central regions and inner disk where the $\ant$ lies in the 
range 0.6 to 1, the fields match within $\sim$10\%. However, in the outer 
parts of these galaxies where $\ant$ is steeper ($>1.2$), the classical 
equipartition values are overestimated by $>20$\% and increases with steepening
of the spectral index to up to 50--60\% towards the edge. Such a deviation
between magnetic fields estimated by the two methods was shown in
\citet{beck05}.  Thus, the magnetic field determined using the classical
formula is found to be constant or increasing as a function of galactocentric
distance. 

\subsection{Comparison with existing studies}

NGC 1097 was studied in polarization at high resolution that revealed magnetic
field in the bar to be aligned with the gas streamlines and thus a good tracer
of gas flow \citep{beck99}. Strong radio emission is detected from the bar at
$\lambda90$ cm \citep{basu12a}, however, due to poor resolution of $\lambda20$
cm maps ($\sim40$ arcsec), the enhancement of magnetic field in the bar is only
about 10--15 percent higher than the disk.  In this study, the field at the
center is found to be $\sim18~\mu$G and decreases to $\sim10~\mu$G towards the
edge. The field in the northern bar is found to be lower than that in the
southern bar with $\langle B_{\rm eq}\rangle \sim 9.8~\mu$G and $\sim
12.2~\mu$G respectively. Our estimated field is lower than what was estimated
by \citet{beck05ea} perhaps due to their assumption of 500 pc of synchrotron
emitting region.

\begin{figure*} 
\includegraphics[width=7cm]{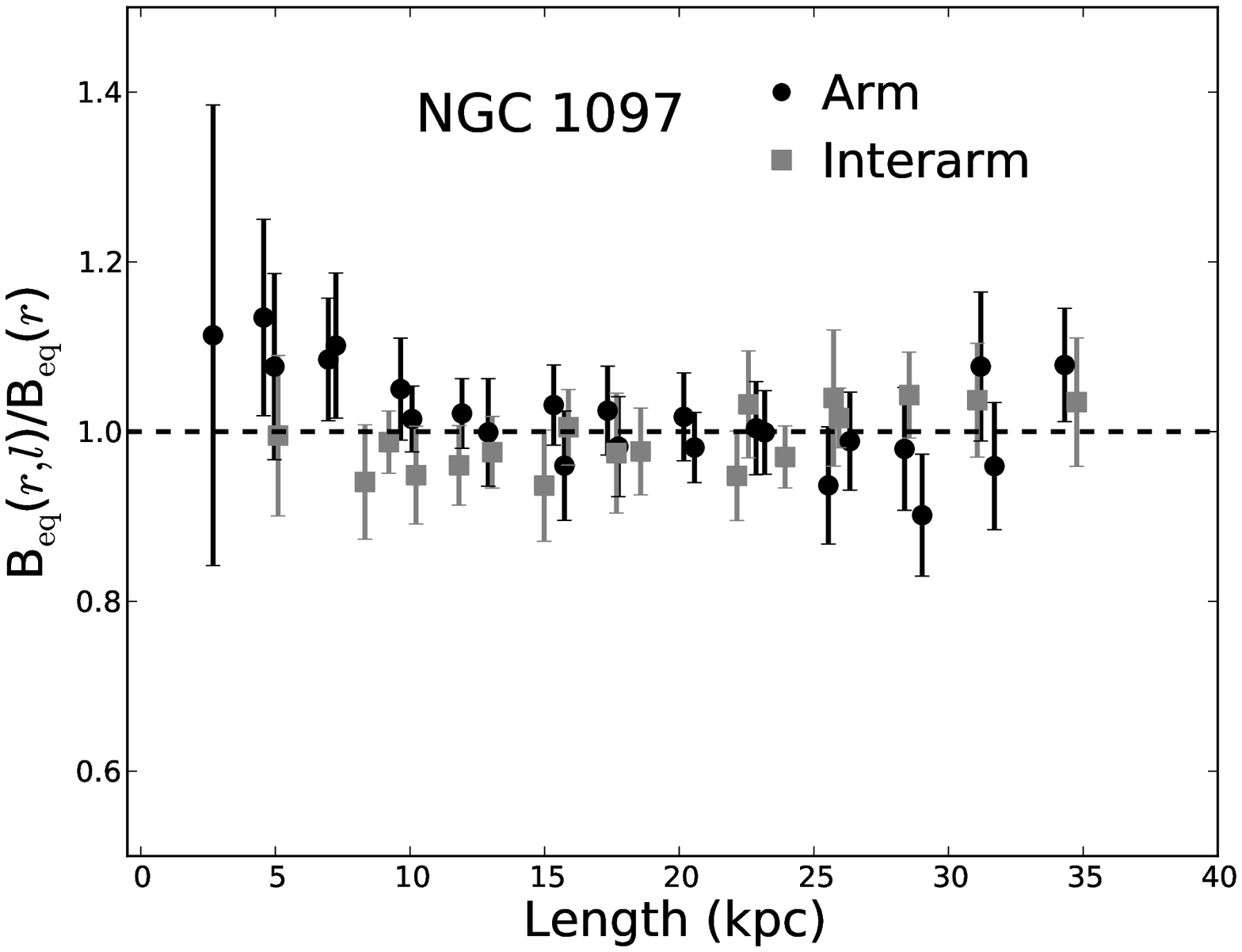} 
\includegraphics[width=7cm]{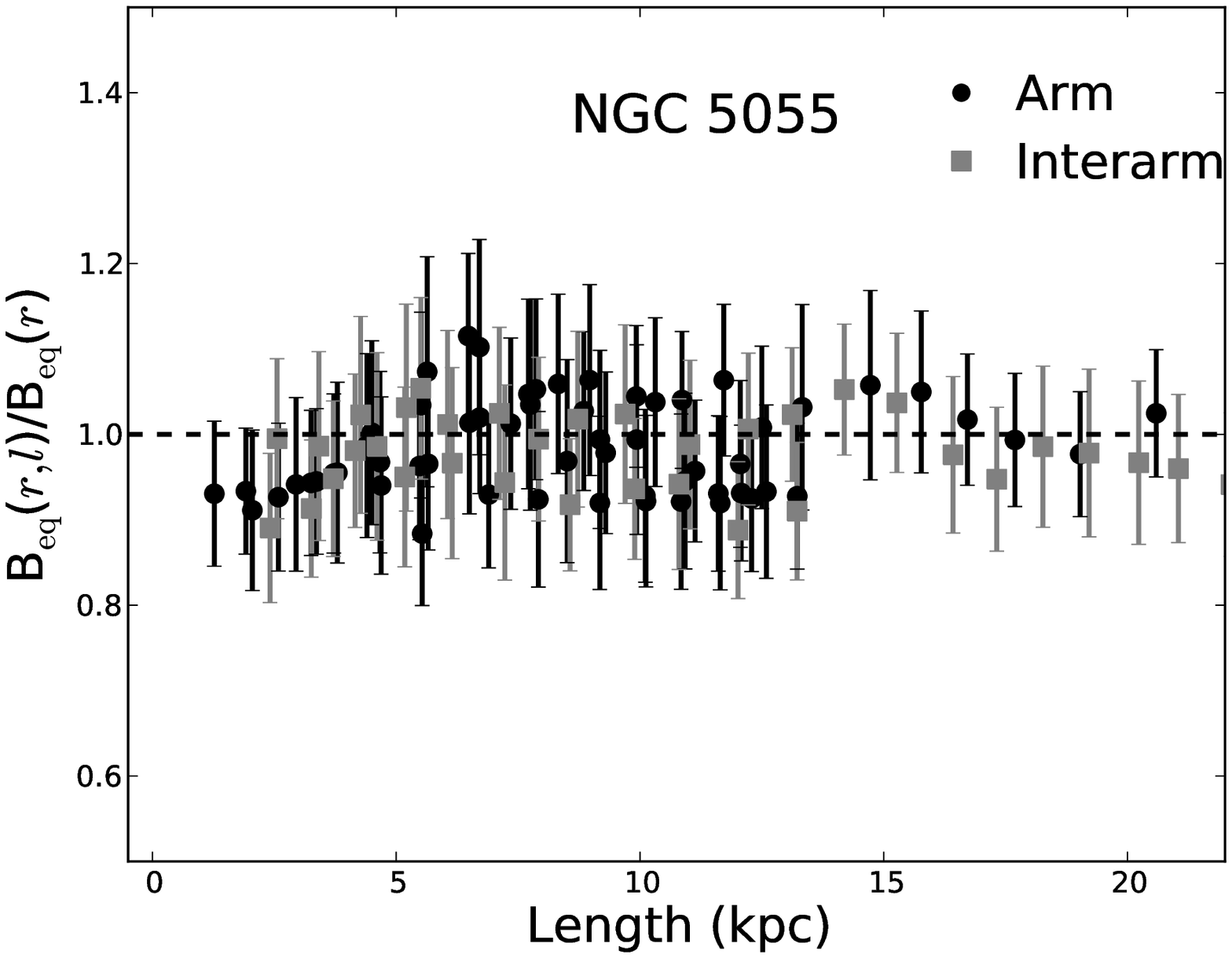} 
\includegraphics[width=7cm]{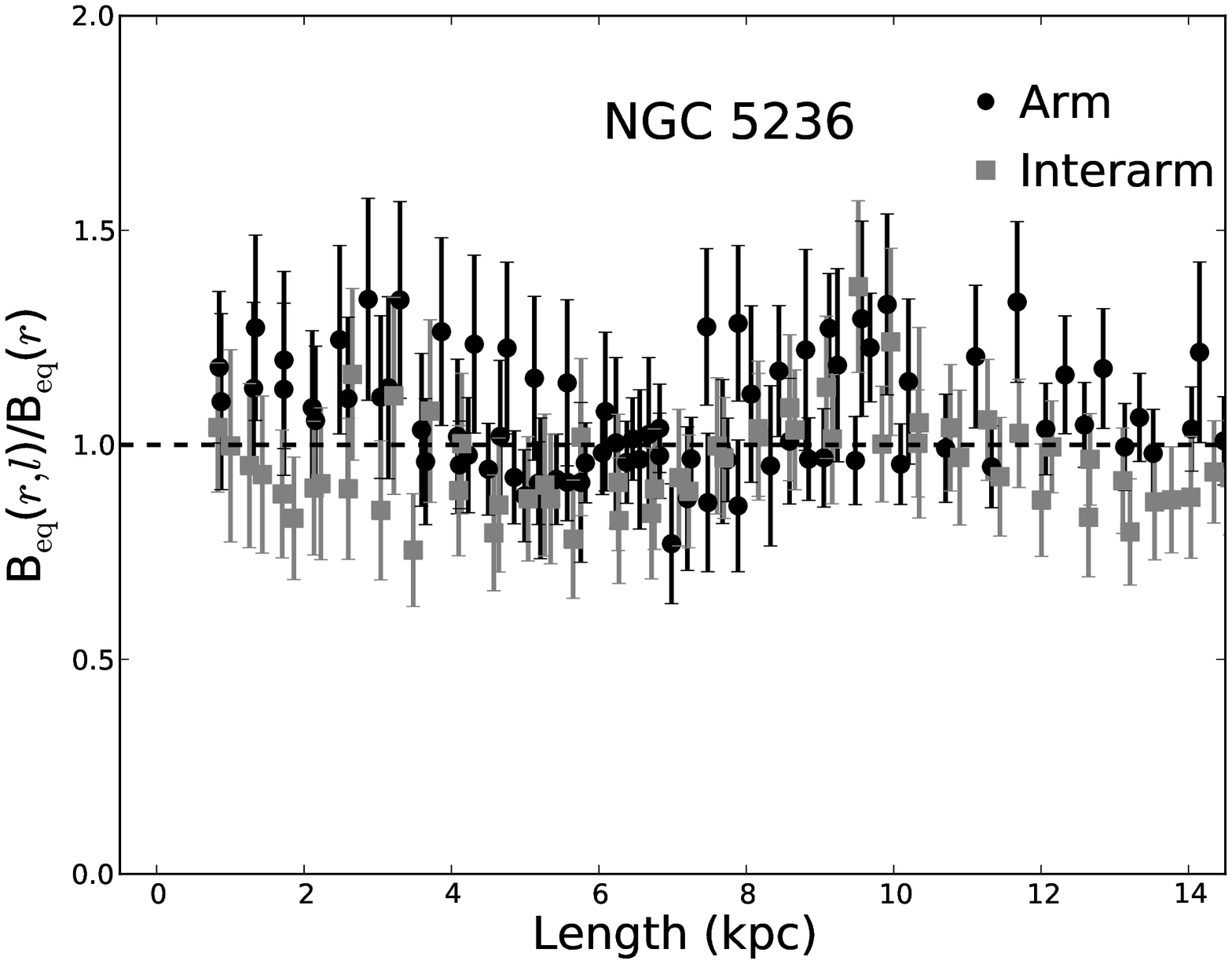} 
\includegraphics[width=7cm]{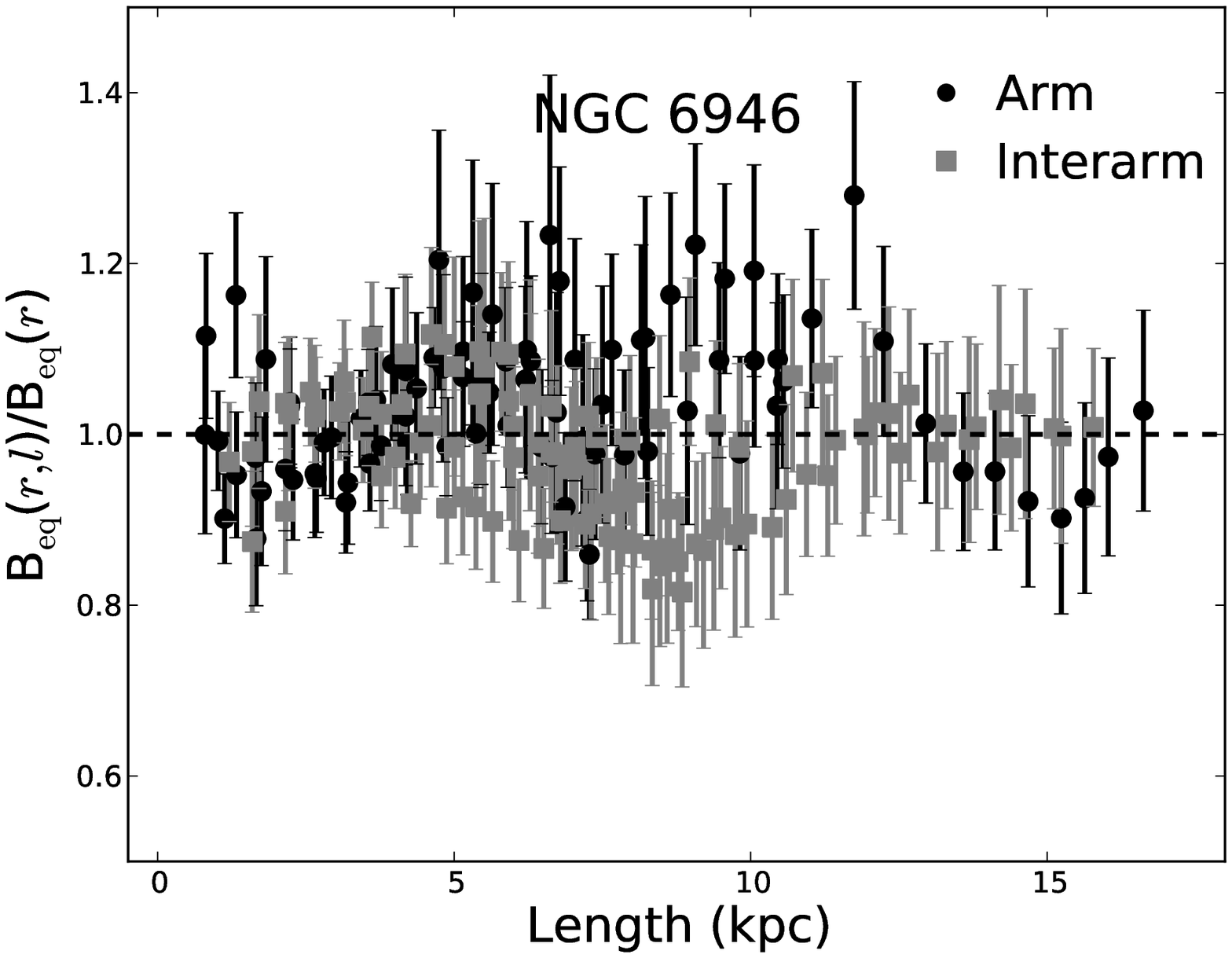} 
\caption{Variation of total magnetic field strength along arm and interarm 
after correcting for the galactocentric variation as in Fig. 2.} 
\label{bratio} 
\end{figure*}

NGC 4736 is a ringed galaxy with no prominent spiral structure from radio
through infrared to optical. Polarized radio emission revealed ordered magnetic
field in spiral shape possibly amplified by large-scale MHD dynamo
\citep{chyzy08}. They report mean total magnetic field of $17~\mu$G, slightly
higher than our estimate of 14 $\mu$G within a region of $\sim3.5$ arcmin
(corresponds to inner 2.3 kpc radius).  From our map (see Fig.~\ref{bmaps}),
the total magnetic field in the center is found to be $\sim18~\mu$G while in
the ring, the field strength is 15--25 $\mu$G with an average total field
strength of $\sim16.5~\mu$G close to \citet{chyzy08} . Beyond the ring the
magnetic field falls off to about 8--10 $\mu$G at a distance of $\sim3$ kpc.

NGC 5055 is a flocculent spiral galaxy and lacks organized spiral structure
when seen in optical. Polarization observations revealed regular spiral
magnetic fields believed to have been generated from turbulent dynamo action
\citep{knapi00}. They estimated a mean equipartition magnetic field of 9.2
$\mu$G close to our value of 9.5 $\mu$G. Of all the galaxies in the sample,
NGC 5055 has the weakest total magnetic field of $\sim14.5~\mu$G in the center
and falls off to about 10 $\mu$G in the disk and $\sim8~\mu$G towards the edge.
No distinct spiral structure has been seen in the map of total magnetic field.

NGC 5236 (M83) is a starburst galaxy with uniform magnetic field seen in the
outer parts of the galaxy and lower degree of uniformity towards the inner
regions hosting star formation \citep{sukum89, neini91, neini93}. Towards the
center and inner spiral arms which harbors the starburst \citep{calze99}, the
$\ant$ was found to be flatter and lies in the range 0.4--0.55. For those
regions we have assumed the value of $\ant$ as 0.55 to avoid any sudden
discontinuities. The magnetic field strengths are overestimated in such
regions. We found a mean total field strength of 24 $\mu$G in the central 1
kpc regions. The magnetic field is found to be strong in the arms with strength
$\sim$15--20 $\mu$G and falls to $\sim10~\mu$G in the interarms and towards the
edge. The mean total field in the galaxy is found to be 12.2$\pm$2.5 $\mu$G,
close to what was estimated by \citet{neini91} within measurement errors.

Magnetic field in NGC 6946 was studied in detail by \citet{beck07}. The total
magnetic field strength was found to be $\sim20~\mu$G in the spiral gas arms,
close to what is estimated by us. In the arms turbulent fields dominate, while
in the interarms large scale regular field was seen with high degree of
polarization (30--60\%) referred as the ``magnetic arms'' by \citet{beck07}.
From our maps (see Fig.~\ref{bmaps}), in the northern (roughly centered at RA =
$\rm 20^h34^m52^s$, DEC = $+60^\circ11^\prime59\arcsec$ J2000) and southern
(roughly centered at RA = $\rm 20^h34^m33^s$, DEC =
$+60^\circ06^\prime46.75\arcsec$ J2000) magnetic arms, the average field is
found to be $\sim11~\mu$G, which is just 10--15\% stronger than other interarm
regions, and this result is at  $\gtrsim$3$\sigma$ significance. The ``circular
hole'' of $\sim1$ kpc diameter seen in the galaxy with low radio emission at RA
= $\rm 20^h34^m20^s$ and Dec. = $+60^\circ09^\prime40\arcsec$ (J2000) is seen
to have low magnetic field ($\sim8.5~\mu$G) as compared to other parts and is
$\sim30\%$ lower than the surrounding regions.

\subsection{Magnetic fields in arms and interarms} 

\begin{table*} 
\begin{centering} 
\caption{Exponential scale lengths ($r_0$) of ISM components. } 
\begin{tabular}{@{}cccccc@{}} 
\hline 
& NGC 1097 & NGC 4736 & NGC 5055 & NGC 5236 & NGC 6946 \\ 
Component& (kpc) &(kpc) & (kpc) &(kpc) & (kpc)\\ 

\hline 
$I_{\rm nt, 20cm}$    & 2.72$\pm$0.28  & 0.95$\pm$0.10 & 3.21$\pm$0.08 & 2.51$\pm$0.32 & 4.05$\pm$0.34\\ 
$I_{\rm nt, 90cm}$    & 4.81$\pm$0.92  & 1.30$\pm$0.26 & 6.17$\pm$0.52 & 4.11$\pm$0.35 & 5.29$\pm$0.42\\ 
$B_{\rm eq}$    & 12.98$\pm$2.93  & 5.0$\pm$0.3  & 12.38$\pm$0.50 & $>$7.74 & 17.70$\pm$1.00\\ 
$\Sigma_{\rm gas}$   & --              & 0.85$\pm$0.04  & 4.25$\pm$0.16  & 2.20$\pm$0.13 & 3.58$\pm$0.22\\ 
\hline 
\end{tabular} 
\end{centering} 
\label{scale} 
\end{table*}

We studied the variation of magnetic field along arm and interarm regions for
the galaxies NGC 1097, NGC 5055, NGC 5236 and NGC 6946 after correcting for the
radial variations. This was not possible for the ringed galaxy NGC 4736, which
do not have any prominent arm.  Arms and interarm regions were chosen using the
Spitzer $\lambda24~\mu$m images after smoothing to the resolution of nonthermal
maps.  For each of the galaxies, the beginning of arm or interarm were chosen
leaving the central $\sim1$ kpc region. We determined magnetic field within an
area of one synthesized beam ensuring no overlap between adjacent beams.  
Each beam corresponds to $\sim$0.4 to 2 kpc at the distance of the galaxies
(see Table 2).  The galaxy NGC 5055, where the arm and interarm are not clearly
distinguishable, the mean field in arm was only about 5 percent stronger than
that in the interarm. For the other galaxies the mean magnetic field in the
arms are stronger by 10--15 percent (see Table~\ref{bmeantab}). Overall, the
mean magnetic field strength in the arm is higher than that in the interarm by
$12\pm3$ percent, however, in certain regions it could be higher by up to 40
percent.  We note that, when magnetic fields are higher in the arms, the
limited telescope resolution suppresses the observed
field strength in the arms and increases that in the interarm regions. The same
effect is caused by larger CRe diffusion length at $\lambda90$ cm, so that more
radio emission is observed in the interarm regions. Thus the differences seen 
in the magnetic field strength between arm and interarm regions are lower limits.

In Figure~\ref{bratio}, we study the variation of the relative magnetic field
strength $B_{\rm eq}(r,d)/B_{\rm eq}(r)$, where, $r$ is the galactocentric
distance and $d$ is the linear distance measured from the center along the
corresponding arm or interarm. In the figure, the black circles and the gray
squares represent arms and interarms respectively. After correcting for the
radial variation, the magnetic field do not change significantly along arm or
interarm. The mean value of $B_{\rm eq}(r,d)/B_{\rm eq}(r)$ in the arm is found
to be $1.03\pm0.03$ and $0.97\pm0.02$ in the interarms for all the galaxies
combined.

\subsection{Radial scale lengths} 

It is believed that CRe originates from supernova explosions of OB stars found
in \hii complexes, i.e., regions of star formation. These CRe then propagate
away to larger distances in galaxies giving rise to larger radial distribution
of synchrotron emission than that of CRe sources and gas \citep[see][]{tabat07,
beck07}.  The total intensity radio maps at $\lambda90$ cm appears to be
significantly smoother than that at $\lambda20$ cm \citep[see][]{basu12a}. The
former mainly originates from older ($\sim10^8$ yr) population of CRe that
diffuses farther away from their formation sites than that at $\lambda20$ cm.
We estimate the exponential scale length of nonthermal emission at $\lambda20$
cm ($I_{\rm nt, 20cm}$) and $\lambda90$ cm ($I_{\rm nt, 90cm}$), total
equipartition magnetic fields and  surface mass density of total gas
($\Sigma_{\rm gas}$) to explore the effect of diffusion of CRe.  $\Sigma_{\rm
gas}$ is computed from atomic and molecular hydrogen surface mass density (see
Appendix A).  The scale lengths ($l_0$) were obtained by fitting a function
$f(r) = f_0 \exp(-r/l_0)$ to the radial profiles shown in left panel of
Fig.~\ref{energy} leaving aside the central bulge. For the ringed galaxy NGC
4736, the scale lengths were computed leaving aside the ring. The scale lengths
obtained are summarized in Table 3. 

\begin{figure*} 
\includegraphics[scale=.35]{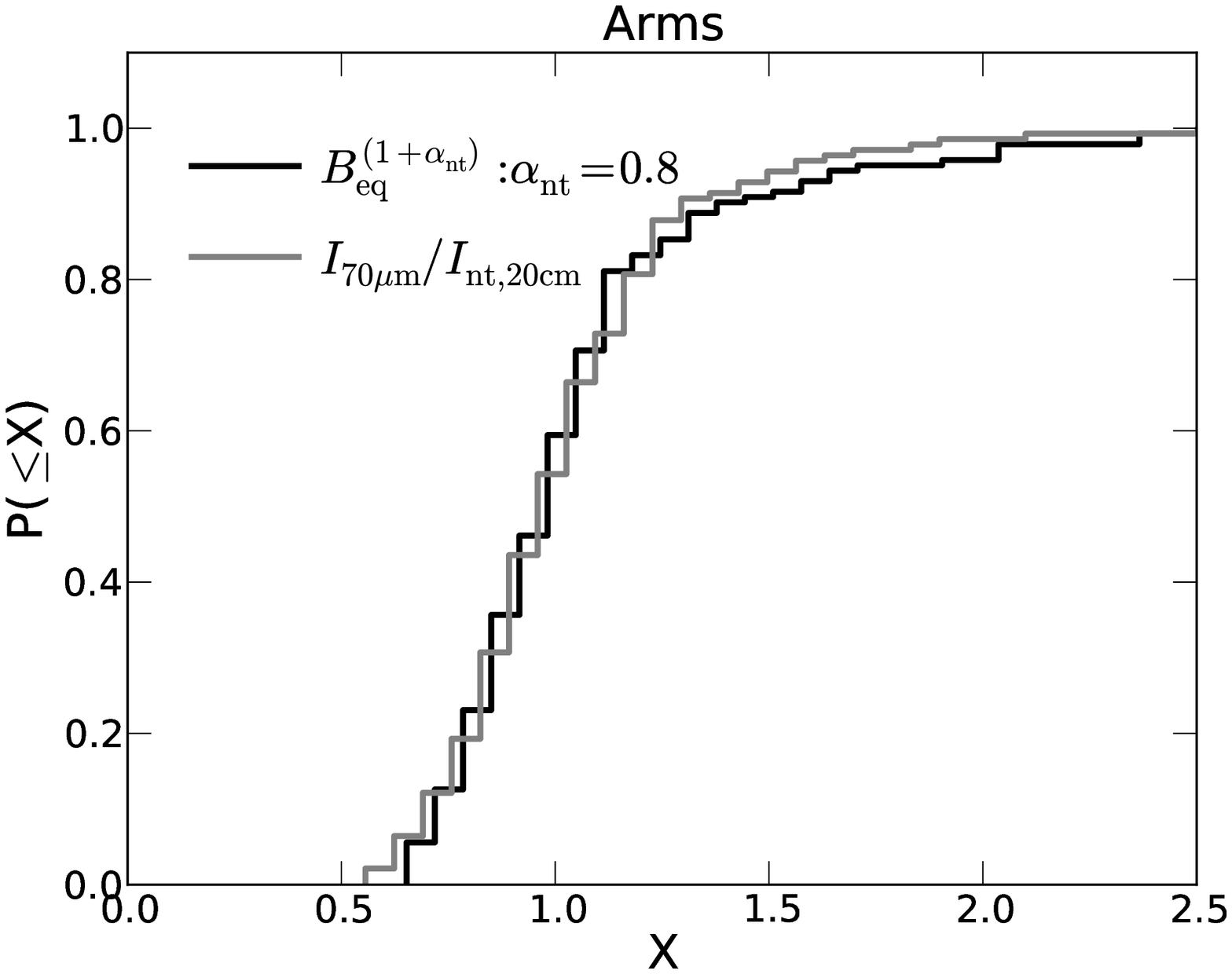} 
\includegraphics[scale=.35]{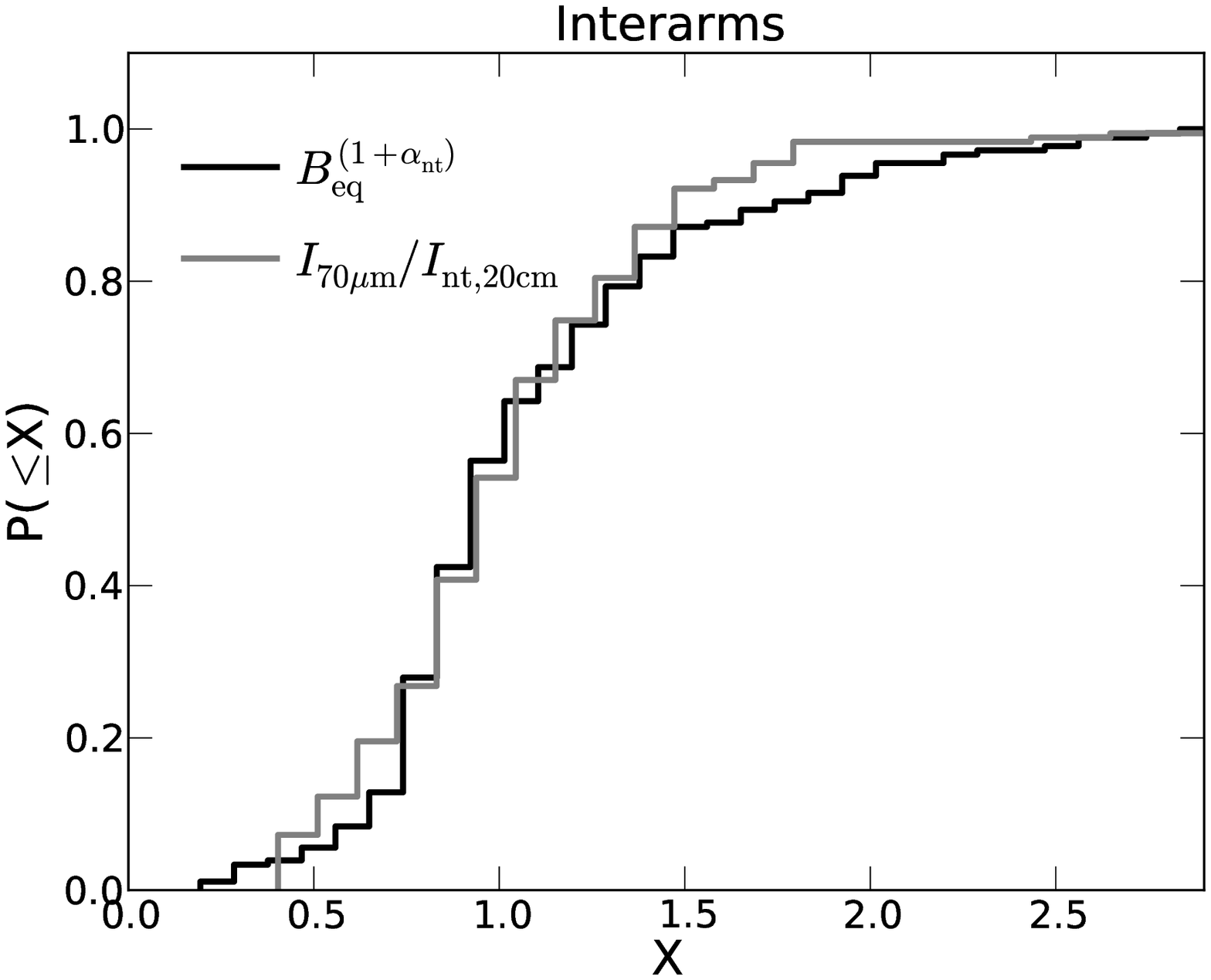} 
\caption{The cumulative distribution function of $X = f_{\rm 70~\mu m}/f_{\rm
20cm}$ (shown in gray) and $X = B_{\rm eq}^{(1+\ant)}$ (shown in black) for arm
(left panel) and interarm (right panel). In the arm we use $\ant=0.8$ while in
the interarm $\ant$ is determined for each of the corresponding region (see text 
for details).} 
\label{cumdistr} 
\end{figure*}

The scale length of the nonthermal emission ($l_{\rm nt}$) at $\lambda90$ cm
was found to be higher than that at $\lambda20$ cm. This is caused due to
higher diffusion scale lengths of low energy ($\sim1.5$ GeV) CRe at $\lambda90$
cm as compared to higher energy ($\sim3$ GeV) CRe at $\lambda20$ cm
\citep{basu12b} in a typical galactic magnetic field of $\sim10~\mu$G. In the
simple case of energy dependent diffusion of CRe, the diffusion length ($l_{\rm
diff}$) after time $\tau$ is given by, $l_{\rm diff} \sim (D~\tau)^{0.5}$.
Here, $D$ is the diffusion coefficient assumed to be constant, and is $\sim
10^{28}~{\rm cm^2s^{-1}}$.  We assume the diffusion time to be same as the
synchrotron cooling timescales ($t_{\rm syn}$) given by, $t_{\rm syn} = 8.35
\times 10^9~\left(E_{\rm CRe}/{\rm GeV}\right)^{-1} \left(B_{\rm eq}/\mu
G\right)^{-2}~{\rm yr}$, where, $E_{\rm CRe}$ is the energy of the electrons.
The expected diffusion length for the galaxies at $\lambda90$ cm and
$\lambda20$ cm are $\sim1.4$ kpc and $\sim1$ kpc respectively, i.e, $l_{\rm
diff,90cm}/ l_{\rm diff,20cm} = (E_{\rm CRe,90cm}/E_{\rm CRe,20cm})^{-0.5}\sim
1.4$. CRe can also propagate by the streaming instability at the
velocity of Alfv$\acute{\rm e}$n wave in the ionized galactic medium and the
propagation distance is given by, $l_{\rm A} = v_{\rm A} t_{\rm syn}$. Here,
$v_{\rm A}$ is the Alfv$\acute{\rm e}$n velocity assumed to be $\sim50$ km
s$^{-1}$. In this scenario, the propagation distance at $\lambda90$ cm and
$\lambda20$ cm are $\sim$1 kpc and $\sim$2 kpc respectively, i.e., $l_{\rm
A,90cm}/l_{\rm A,20cm}= (E_{\rm CRe,90cm}/E_{\rm CRe,20cm})^{-1} \sim 2$.  From
our data, the ratio of scale length of nonthermal emission at $\lambda90$ cm
and $\lambda20$ cm, i.e, $l_{\rm nt,90cm}/l_{\rm nt,20cm}$ are $1.77\pm0.38$,
$1.37\pm0.29$, $1.92\pm0.17$, $1.64\pm0.25$ and $1.31\pm0.15$ for NGC 1097, NGC
4736, NGC 5055, NGC 5236 and NGC 6946 respectively.  For 3 of the galaxies the
increase in the estimated $l_{\rm nt}$ between $\lambda90$ and $\lambda20$ cm
is larger than that expected from simple diffusion estimates and is consistent
with streaming with Alfv$\acute{\rm e}$nic velocity. 

The scale length of nonthermal emission at $\lambda20$ cm for NGC 6946 is
similar to what was found by \citet{walsh02} and \citet{beck07}.  The
nonthermal scale length is related to scale length of magnetic field ($l_B$) as
$l_B = l_{\rm nt} (3 + \ant)$ under the assumption of equipartition of
energy between magnetic field and cosmic ray particles.  For the galaxy NGC
5236, $l_B$  is found to be comparatively smaller than other galaxies and is
only $\sim3$ times than that of $l_{\rm nt}$ at $\lambda20$ cm and $\sim$1.9
times at $\lambda90$ cm.  This is perhaps the effect of magnetic field strength
being overestimated towards the inner parts of the galaxy, where $\ant \leq
0.5$ (see Section~2.1).  Thus the estimated $l_B$ for NGC 5236 is lower than
the actual value. The scale length of the magnetic field for NGC 6946 is found
to be slightly higher than what was estimated by \citet{beck07}. This is likely
to be caused due to their assumption of a constant $\ant$ throughout the galaxy
and use of nonthermal emission at $\lambda20$ cm which has a smaller $l_{\rm
nt}$ as compared to our $\lambda90$ cm maps.

The scale length of total gas surface density ($l_{\rm gas}$) is found to be
smaller than that of the nonthermal emission. However, $l_{\rm gas}$ is close
to $l_{\rm nt}$ at $\lambda 20$ cm and much smaller than $l_{\rm nt}$ at
$\lambda90$ cm, suggesting the $\lambda20$ cm nonthermal emission is a better
tracer of star forming activity than at $\lambda90$ cm, wherein the later
mostly traces the older population of CRe which are well mixed.

\section{Discussions} 

\subsection{Is synchrotron intensity an indicator of magnetic field ?} 

\subsubsection{Slope of the radio--FIR correlation} 

Simulations of MHD turbulence in the ISM revealed, under conditions of
equipartition, the magnetic field ($B$) and the gas density ($\rho_{\rm gas}$)
are coupled as $B \propto \rho_{\rm gas}^{\kappa}$, where $\kappa\sim0.4-0.6$
\citep[see e.g.,][]{fiedl93, grove03}.  The slope of the well known radio--far
infrared (FIR) correlation was used to determine $\kappa$ for four of the
galaxies at scales of $\sim1$ kpc \citep[][]{basu12b} using synchrotron
emission at $\lambda90$ cm and $\lambda20$ cm and FIR emission at
$\lambda70~\mu$m.  The estimated value of $\kappa$ was found to be
$0.51\pm0.12$, indicating energy `equipartition' among magnetic field and 
kinetic energy of gas due to turbulent motions.  However, in this method
equipartition between magnetic field and cosmic ray particles is assumed
a-priori. The validity of this assumption can be checked from the dispersion
seen in the radio--FIR correlation and our estimated values of the magnetic
fields.

\begin{figure*} 
\includegraphics[width=8.5cm, height=12cm]{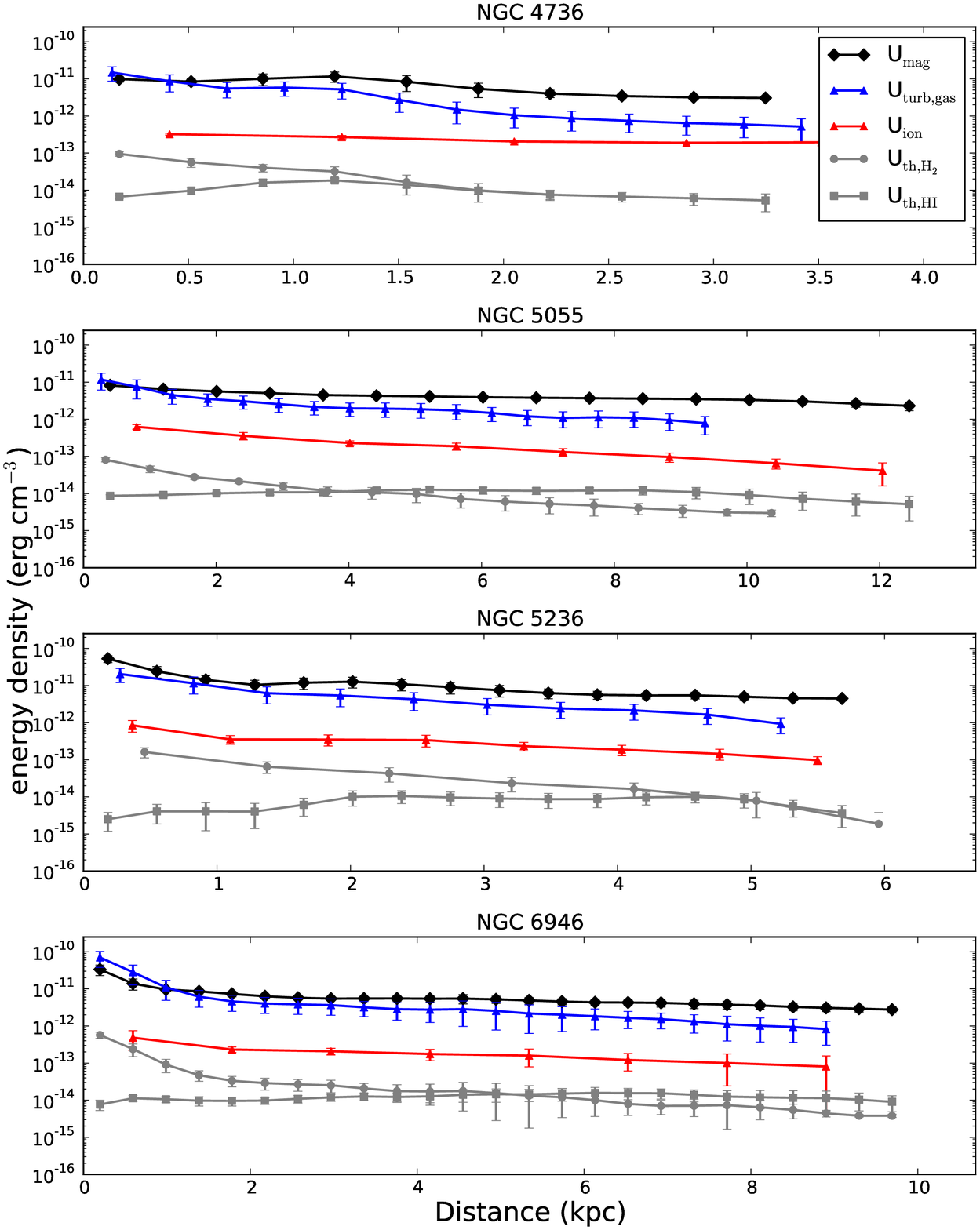} 
\includegraphics[width=8.5cm, height=12cm]{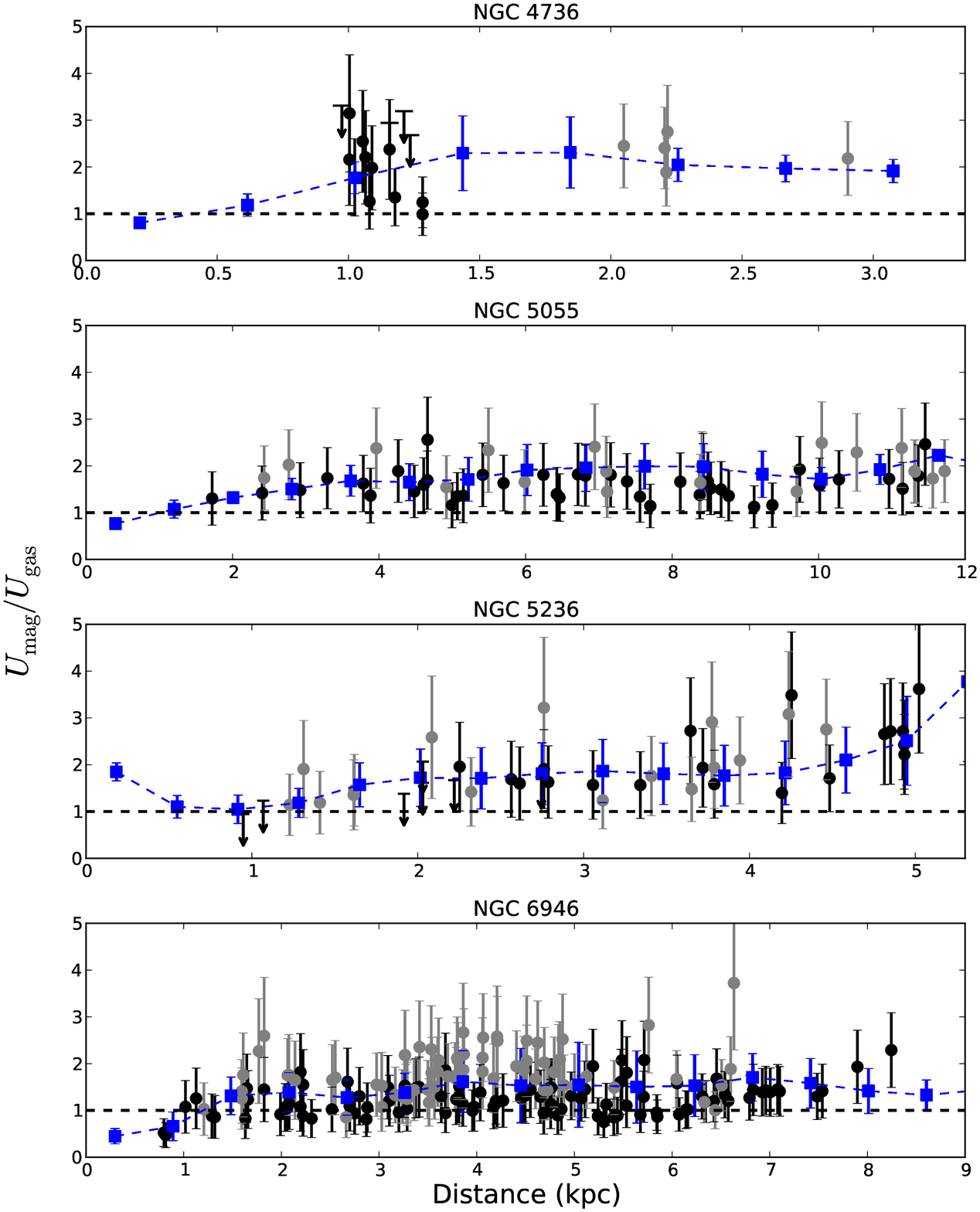} 
\caption{Left panel: energy densities in various ISM phases as a function of
galactocentric distance. Black lines shows the magnetic field energy density
($U_{\rm mag}$), the blue lines shows the kinetic energy density of total
neutral gas due to turbulent motion ($U_{\rm turb,gas}$), the red lines shows
the thermal energy density of warm ionized gas ($U_{\rm ion}$), the gray
lines with dots and squares shows the thermal energy density of atomic and
molecular gas ($U_{\rm th,HI}$ and $U_{\rm th,H_2}$) respectively.  Right
panel: the ratio of energy density in magnetic field and total ISM gas energy
density ($U_{\rm gas} = U_{\rm turb,gas} + U_{\rm ion} + U_{\rm th,neutral}$) at
scales of $\lesssim1$ kpc as a function of galactocentric distance. The black and
gray symbols are for arm and interarm regions respectively. The blue squares
shows the radially averaged value of the ratio determined within annulus of one
one synthesized beam width.}
\label{energy} 
\end{figure*} 

\subsubsection{Dispersion of the radio--FIR correlation} 
\label{sec-disp} 

Dispersion of quantity `$q$' defined as $\log_{10}(I_{\rm FIR}/I_{\rm nt,\nu})$
is widely used as a measure of the tightness of the radio--FIR correlation.
Where, $I_{\rm FIR}$ is the FIR flux density and $I_{\rm nt,\nu}$ is the
nonthermal radio flux density at frequency, $\nu$. The far infrared flux
density ($I_{\rm FIR}$) can be written as, $I_{\rm FIR} \propto n_{\rm UV}
Q(\lambda, a) B(T_{\rm dust})$, where $n_{\rm UV}$ is the number density of the
UV photons responsible for dust heating, $Q(\lambda, a)$ is a
wavelength (here $\lambda=70~\mu$m) dependent absorption coefficient
of dust grains of radius $a$ \citep{drain84, alton04}.  $B(T_{\rm dust})$ is
the Planck function for dust emitting at temperature $T_{\rm dust}$.  The flux
density at radio frequency $\nu$, can be written as $I_{\rm nt,\nu} \propto
n_{\rm CRe,\nu} B^{1+\ant}$, where, $n_{\rm CRe,\nu}$ is the number density of
cosmic ray electrons (CRe) emitting at a frequency $\nu$ and $B$ is the actual
magnetic field.  From the above, the ratio of FIR and radio flux density is,
$I_{\rm FIR}/I_{\rm nt,\nu} \propto (n_{\rm UV}/n_{\rm
CRe,\nu})(1/B^{1+\ant})$, assuming same dust properties throughout the
galaxy and $T_{\rm dust}$ is seen to remain constant throughout the galaxy
\citep[see e.g.][]{tabat07, basu12a}. \citet{humme86} showed that the
cumulative frequency distribution of $I_{\rm FIR}/I_{\rm nt,\nu}$ and that of
$B_{\rm eq}^{1+\ant}$ follows each other indicating energy `equipartition'
between magnetic field and cosmic ray particles to hold good and thus $B_{\rm
eq}$ is close to $B$.  In our case, for each of the galaxy, we determine
the quantities $I_{\rm 70\mu m}/I_{\rm nt,20cm}$ and $B_{\rm eq}^{1+\ant}$
within a region of $\sim1$ kpc and normalized them with their respective median
values.  In Fig.~\ref{cumdistr} we plot these median normalized cumulative
distribution function of $I_{\rm 70\mu m}/I_{\rm nt,20cm}$ (shown in gray)
and $B_{\rm eq}^{1+\ant}$ (shown in black) for all the galaxies together. The
left and right panels show the distribution in the arm and interarm regions
respectively. In the arm regions, $\ant$ do not change significantly, we
assumed a constant value of $0.8$. However, $\ant$ varies significantly in the
interarm regions, and we have used the observed values of $\ant$ for each
region from \citet{basu12a}. 

At $\lambda20$ cm, the dispersion in the quantity $I_{\rm 70\mu m}/I_{\rm
nt,20cm}$ is similar to the dispersion in $B_{\rm eq}^{1+\ant}$ for both arms
and interarm regions determined at spatial scales of $\sim1$ kpc.  Thus at
scales of $\sim$1 kpc, the variations in $I_{\rm 70\mu m}/I_{\rm
nt,20cm}$, i.e., dispersion seen in the quantity `q' is caused
due to variations in the magnetic field, where the magnetic field is
represented by $B_{\rm eq}$. Thus, $B_{\rm eq}$ (or a constant multiple of it)
is a good representative of the actual magnetic field, $B$.

However, at $\lambda90$ cm, the dispersion in $I_{\rm 70\mu m}/I_{\rm nt,90cm}$
is $\sim$20 percent higher than that of $B_{\rm eq}^{1+\ant}$ for the interarm
regions.  At $\lambda90$ cm the low energy ($\sim1.5$ GeV) CRe propagate to
farther distances from the arms into the interarms, which has the effect of
increasing the dispersion.  We note that equipartition assumption is
valid only at scales larger than the diffusion length, which is better 
fulfilled at $\lambda20$ cm than at $\lambda90$ cm.

\subsection{Energy density in magnetic field and gas} 

Magnetic energy is expected to be in equipartition with ISM turbulent energy
\citep{ crutc99, cho00, grove03}. In Section 3, we found that the magnetic
field falls off as a function of galactocentric distance and had a larger scale
length than that of the gas surface density. Here, we compare the magnetic
field energy density ($U_{\rm mag} = B_{\rm eq}^2/8\pi$) with that of the ISM
energy density from kinetic energy of gas due to turbulent motions ($U_{\rm
turb,gas}$), thermal energy density of warm ionized gas ($U_{\rm ion}$) and
total neutral (atomic; $U_{\rm th,HI}$ + molecular; $U_{\rm th,H_2}$) gas at
spatial scales of 0.4 -- 0.9 kpc, except for NGC 5236 for which the spatial
resolution is $\sim1.2$ kpc (see Appendix A for details).  $U_{\rm turb,gas}$
is estimated from the surface mass density maps of atomic and molecular
hydrogen, using $U_{\rm turb,gas}= 1.36(U_{\rm turb,HI} + U_{\rm turb,H_2})$.
The factor 1.36 is to account for the presence of Helium and $U_{\rm turb,HI,
H_2} = (1/2) (\Sigma_{\rm HI, H_2}/h_{\rm HI, H_2}) v_{\rm turb}^2$, where,
$\Sigma_{\rm HI, H_2}$ are the surface mass density of atomic (H{\sc i}) and
molecular (H$_2$) gas. $v_{\rm turb}$ is the velocity of the turbulent gas,
assumed to be $\sim9~\rm km~s^{-1}$ for H{\sc i} and $\sim 6~ \rm km~s^{-1}$ for
H$_2$ \citep{kruit82, combe97, sellw99, kaspa08} and $h_{\rm HI, H_2}$ are the
line of sight depth of atomic and molecular gas assumed to be $\sim400$ pc and
$\sim 300$ pc respectively.  The surface mass densities were calculated using
moment-0 maps of CO and H{\sc i} line emission (see Appendix A for details). 

The thermal energy densities of warm ionized gas and neutral gas were
computed using, $U_{\rm th} = \frac{3}{2} \langle n \rangle k T$. Here,
$\langle n\rangle$ is the mean number density, $k$ is the Boltzmann constant
and $T$ is the temperature. For the warm ionized gas, the mean number density
of thermal electrons $\langle n_{\rm e}\rangle$ was calculated from the
emission measure ($EM$) maps, such that, $\langle n_{\rm e}\rangle \approx
\left[ EM~f_{\rm d} /h_{\rm ion}\right]^{1/2}$.  $EM$ was determined from
dereddened H$\alpha$ maps using Equation 9 in \citet{valls98} \citep[see][for
details]{basu12a}.  We assumed a constant filling factor ($f_{\rm d}$) of
$\sim$5 percent and scale height ($h_{\rm ion}$) of the ionized medium as 1 kpc
\citep{wang97, hoope99}. The temperature, $T_{e}$ was assumed to be $10^4$ K.
To estimate the energy density of molecular ($U_{\rm th,H_2}$) and atomic
($U_{\rm th,HI}$) gas, the number densities were determined from the corresponding
surface mass density maps assuming a constant scale height as discussed above.
We assumed a constant temperature of $\sim50$ K for molecular gas and $\sim100$
K for atomic gas.

In Figure~\ref{energy} (left panel) we study the variation of energy density of
magnetic field ($U_{\rm mag}$; shown in black diamonds), kinetic energy of
total gas ($U_{\rm turb,gas}$; shown in blue triangles) and thermal energy
density of ionized gas ($U_{\rm ion}$; shown in red triangles), molecular gas
($U_{\rm th,H_2}$; shown in gray circles) and atomic gas ($U_{\rm th,HI}$; shown
in gray squares) as a function of galactocentric distance for the galaxies NGC
4736, NGC 5055, NGC 5236 and NGC 6946. The energy density of the warm
ionized gas is found to be about two orders of magnitude lower than that of
magnetic field and kinetic energy of turbulent gas. For neutral gas, the
thermal energies are about 3 orders of magnitude less. The energy density of
magnetic field, turbulent energy of total gas, thermal energy of warm ionized
gas and neutral gas matches well with \citet{beck07} for the galaxy NGC 6946.
However, due to simplistic assumption for $\ant \sim 1.0$ throughout the
galaxy, the total magnetic field may be underestimated in \citet{beck07},
specially in the inner regions where $\ant$ is smaller than 0.7. 

In Figure~\ref{energy} (right panel) we study the variation of the ratio of
magnetic field energy density and total ISM gas energy density ($U_{\rm gas}$)
estimated at spatial scales of $\sim$0.4--1.2 kpc depending on the distance of
the galaxies.  $U_{\rm gas}$ was computed as, $U_{\rm gas} = U_{\rm
turb,gas} + U_{\rm ion} + 1.36(U_{\rm th,HI} + 2U_{\rm th,H2})$. The black and
gray symbols denote arms and interarms respectively.  The blue squares show the
radially averaged value of the ratio determined within annulus of one beam
width. Pixels more than 4$\sigma$ in gas density were considered for this
analysis. 

The ratio of total ISM gas energy density and magnetic field are found to be almost
constant throughout the galaxies except for a systematic offset from unity due
to our assumed constant values of scale heights and velocity dispersion. After
dividing by the mean values of the ratio for each of the galaxies, the ratio
has a dispersion less than $\sim30\%$ for arms and interarms of the galaxies,
indicating energy/pressure balance between magnetic field and gas.  For the
galaxy NGC 6946, we observe that the ratio of the interarm regions to be higher
than that in the arms by a factor $1.47\pm0.11$, i.e, magnetic field energy
dominates over the turbulent gas energy. Overall, this difference is not
significant for other galaxies, but for an adjacent arm-interarm, the ratio in
the interarm could be higher by more than $50\%$ than the arm. 

From our data we found that as we move from the center to the disk to the edge
of the galaxies, magnetic field energy systematically dominates over the 
total ISM energy density.  However, this trend disappears once we consider
only the high signal-to-noise regions ($>4\sigma$) as is shown by the blue
points in Fig.~\ref{energy} (right panel). 

Magnetic fields in galaxies are amplified by two major processes, firstly the
small scale fields are amplified by field line stretching and twisting due to
turbulent motions of gas (small-scale dynamo) and secondly, large-scale
amplification due to large-scale dynamo action. In the previous case,
magnetic field and gas are closely coupled such that the magnetic field is
amplified by transfer of kinetic energy from gas. The fields can be amplified
up to equipartition levels \citep{grove03, cho00}. This would result in small
scale turbulent fields with low degree of polarization in regions of high gas
density, i.e, the spiral arms. This is indeed observed in many of the galaxies.
In NGC 6946 only 1--5\% of the emission is polarized in the inner spiral arms
\citep{beck07}.  In M51, turbulent magnetic field dominates in arms
\citep{fletc11}. In NGC 5236, low degree of uniform magnetic field is seen in
the spiral arms in inner parts of the galaxy \citep{neini93}.  In NGC 4736,
comparatively lower degree of polarization is seen in the star forming ring
than in the outer parts \citep{chyzy08}. Our results show that in the gaseous
arms, the magnetic field energy density and the energy density in gas to be
similar and do not vary by more than 30 percent throughout the galaxies. This
indicates that in the arms the magnetic field is perhaps amplified by field
line stretching due to turbulent gas motions driven by star formation.

In the interarms and towards the outer parts, many galaxies show higher
degree of ordered magnetic field caused by large scale dynamo action
\citep[see e.g.][and references therein]{beck96a, beck07, chyzy08}. For NGC
6946, where ordered fields are observed in the interarm regions, it is thought
that finite time dynamo relaxation causes a phase shift between magnetic and
gas/star forming spiral arms, such that magnetic arms lags \citep{chama13}.
Magnetic field energy dominating over the turbulent gas energy thus helps
maintaining this field orderness in interarm regions and outer parts of the
galaxies.  For this galaxy, \citet{walsh02} found the regular magnetic field
(using polarized emission at $\lambda 6$ cm) to trace regions of low star
formation efficiency and coincides with the regions where the ratio was found
to be significantly higher, suggesting insufficient energy in turbulent gas to
amplify the turbulent magnetic field.

\section{summary} 

We have measured total magnetic fields in five nearby normal galaxies, NGC
1097, NGC 4736, NGC 5055, NGC 5236 and NGC 6946, assuming equipartition of
energy between cosmic ray particles and magnetic fields. In this study, magnetic
fields were probed at sub-kpc scales except for NGC 1097, for that it was 2.8
kpc.

\begin{itemize}

\item The strengths of the total magnetic field decreases by $\sim40-50\%$
from center to edge of the galaxies. The field changes by at least $15\%$
between arms and interarms and do not change significantly along them after
correcting for the radial variation. 

\item Our study shows synchrotron intensity to be a good tracer of the total
magnetic field in galaxies. `Equipartition' of energy between magnetic field
and cosmic ray particles hold well at kpc scales for all the galaxies. 

\item The estimated energy densities of magnetic field and gas were seen to be
within a factor $\lesssim$2 in the arms and interarms at sub-kpc scales
implying magnetic field to play important role in pressure balance of the ISM.
The ratio $U_{\rm mag}/U_{\rm gas}$ is found to be roughly constant along
radius.

\item The energy density of the magnetic field was found to be larger than
that of the kinetic energy density due to turbulent motion of gas in the
interarm regions, particularly for NGC 6946, and in outer parts in general.
Large scale dynamo action could maintain the magnetic field in such regions.

\end{itemize}

\section*{Acknowledgments} 
 We thank the referee Rainer Beck for important comments which considerably 
improved the presentation of the paper. We thank Wilfred Walsh for providing us
the CO$_{\rm J=3\to2}$ moment-2 map for NGC 6946. We would like to thank Adam
Leroy for providing us the FITS files of the moment-0 CO$_{\rm J=2\to1}$ maps.
We thank Dipanjan Mitra for useful discussions and Visweshwar Ram Marthi
for going through the manuscript.  This research has made use of the NASA/IPAC
Extragalactic Database (NED), which is operated by the Jet Propulsion
Laboratory, California Institute of Technology, under contract with the
National Aeronautics and Space Administration.  This work is based (in part) on
observations made with the {\it Spitzer Space Telescope}, which is operated by
the Jet Propulsion Laboratory, California Institute of Technology under a
contract with NASA.

\appendix 
\section{Atomic and Molecular surface gas density} 

Four of the galaxies (NGC 4736, NGC 5055, NGC 5236 and NGC 6946) studied in 
this work were observed as a part of THINGS \citep{walte08} to trace H{\sc i}. 
We used natural weighted moment-0 H{\sc i} maps to calculate the surface 
density of atomic gas using the equation, $$\Sigma_{\rm HI} ({\rm M_\odot 
pc^{-2}}) = 0.02~\cos i~ I_{\rm HI} ({\rm K~km~s^{-1}}).$$ Here, $i$ is the 
inclination angle ($0^\circ$ for face-on) of the galaxy and $I_{\rm HI}$ is the 
line integrated intensity.  The above equation includes a factor 1.36 to 
account for the presence of Helium. 

CO is commonly used as a tracer for molecular gas. We used CO$_{\rm J=2\to1}$
moment-0 maps from the HERACLES \citep{leroy09} for three of the galaxies,
namely NGC 4736, NGC 5055 and NGC 6946. These maps has an angular resolution of
13.4 arcsec, better than the resolution of the nonthermal maps. Assuming a
constant CO-to-H$_2$ conversion factor, $X_{\rm CO} = 2\times10^{20} {\rm
cm^{-2} (K~km~s^{-1})^{-1}}$ and a line ratio of 0.8 for converting CO$_{\rm
J=2\to1}$ flux to CO$_{\rm J=1\to0}$ flux, the molecular gas surface density
was calculated using, $$ \Sigma_{\rm H_2} ({\rm M_\odot pc^{-2}}) = 5.5~\cos
i~I_{\rm CO_{\rm J=2\to1}} ({\rm K~km~s^{-1}}).$$ Here, $I_{\rm CO_{\rm
J=2\to1}}$ is the line integrated intensity. Note that, for NGC 6946, the
$\cos~i$ factor was missing in \citet{beck07} by mistake (Rainer Beck, private
communication). However, this would not change their conclusions
significantly.

For the galaxy NGC 5236 we used CO$_{\rm J=1\to0}$ moment-0 map from the NRAO
12-m telescope \citep{crost02} to calculate the molecular gas density.   This
map has an angular resolution of 55 arcsec.  The line integrated flux density
($S_{\rm CO_{\rm J=1\to0}}$) was converted into molecular gas mass ($M_{\rm
H_2}$) using the formula $$M_{\rm H_2} (M_\odot) = 1.1\times10^4 D({\rm Mpc})^2
\cos i~ S_{\rm CO_{\rm J=1\to0}} ({\rm Jy~km~s^{-1}})$$ from \citet{young89}.
Here, $D$ is the distance to the galaxy.  The mass was then converted to
surface density by diving by the linear area for each pixel. 

\label{lastpage} 

\end{document}